\documentclass[prd,amsmath,amssymb,showpacs,superscriptaddress,nofootinbib,twocolumn]{revtex4-1}
\usepackage{graphicx}
\usepackage{epsfig,graphics,subfigure,psfrag,amssymb}
\usepackage{lineno}
\usepackage{dcolumn}
\usepackage{bm}
\usepackage{overpic}
\usepackage{xspace}
\usepackage{rotating}
\usepackage{color}
\usepackage{multirow}
\usepackage{colortbl}
\usepackage{epstopdf}
\usepackage[colorlinks,linkcolor=blue,anchorcolor=blue,citecolor=blue,urlcolor=blue]{hyperref}
\newcommand{\ra}{\rightarrow}

\newcommand{\BESIII}{BES\uppercase\expandafter{\romannumeral3}\xspace}

\begin{document}
\title{%
{\boldmath
Measurements of $e^+e^- \to K_{S}^{0}K^{\pm}\pi^{\mp}\pi^0$ and $K_{S}^{0}K^{\pm}\pi^{\mp}\eta$ at center-of-mass energies from $3.90$ to $4.60~\mathrm{GeV}$}}

\author{
M.~Ablikim$^{1}$, M.~N.~Achasov$^{10,d}$, S. ~Ahmed$^{15}$, M.~Albrecht$^{4}$, M.~Alekseev$^{55A,55C}$, A.~Amoroso$^{55A,55C}$, F.~F.~An$^{1}$, Q.~An$^{52,42}$, J.~Z.~Bai$^{1}$, Y.~Bai$^{41}$
% ...
\\
\vspace{0.2cm}
(BESIII Collaboration)\\
\vspace{0.2cm} {\it
$^{1}$ Institute of High Energy Physics, Beijing 100049, People's Republic of China\\
% ...
$^{4}$ Bochum Ruhr-University, D-44780 Bochum, Germany\\
% ...
$^{10}$ G.I. Budker Institute of Nuclear Physics SB RAS (BINP), Novosibirsk 630090, Russia\\
% ...
$^{15}$ Helmholtz Institute Mainz, Johann-Joachim-Becher-Weg 45, D-55099 Mainz, Germany\\
% ...
$^{41}$ Southeast University, Nanjing 211100, People's Republic of China\\
$^{42}$ State Key Laboratory of Particle Detection and Electronics, Beijing 100049, Hefei 230026, People's Republic of China\\
% ...
$^{52}$ University of Science and Technology of China, Hefei 230026, People's Republic of China\\
% ...
$^{55}$ (A)University of Turin, I-10125, Turin, Italy; (B)University of Eastern Piedmont, I-15121, Alessandria, Italy; (C)INFN, I-10125, Turin, Italy\\
% ...
$^{d}$ Also at the Novosibirsk State University, Novosibirsk, 630090, Russia\\
% ...
}}
%\date{\today}

\begin{abstract}
 Using $5.2 \ \mathrm{fb}^{-1}$ $e^+ e^-$ annihilation data samples collected with the BESIII detector, we measure the cross
 sections of $e^+e^- \to K_S^0 K^\pm \pi^\mp \pi^0$ and $K_{S}^{0}K^{\pm}\pi^{\mp}\eta$ at center-of-mass energies from
 $3.90$ to $4.60$ GeV. In addition, we search for the charmonium-like resonance $Y(4260)$ decays into $K_{S}^{0}K^{\pm}\pi^{\mp}\pi^0$ and
 $K_{S}^{0}K^{\pm}\pi^{\mp}\eta$, and $Z_c^{0,\pm}(3900)$ decays into $K_{S}^{0}K^{\pm}\pi^{\mp,0}$ and $K_{S}^{0}K^{\pm}\eta$.
 Corresponding upper limits are provided since no clear signal is observed.
\end{abstract}


\maketitle

\section{Introduction}
With the experimental progress in the past decade, many charmonium-like $(XYZ)$ states were observed, which can not be accommodated within the
naive quark model and are proposed
as the candidate of  the hidden-charm exotic mesons~\cite{XYZ_review1,XYZ_review2}.
In this paper, we focus on the exotic states $Y(4260)$ and $Z_c(3900)$.

$Y(4260)$ was first observed in ISR process $e^+ e^- \to \gamma_{\rm{ISR}} \pi^+ \pi^- J/\psi$ by $BABAR$~\cite{ PhysRevLett.95.142001}. In later
experiments, $Y(4260)$  was also observed in $Y(4260)\to\pi^0\pi^0J/\psi$~\cite{Y4260_pi0jpsi}. Recently, BESIII has observed a resonance
around $4.23~\mathrm{GeV}$ in a open-charm process $e^+ e^- \to \pi^+ D^0D^{\star-}$~\cite{Y4260_piddstar}.
But no other open
charm~\cite{Y4260notseen1,Y4260notseen2,Y4260notseen3,Y4260notseen4,Y4260notseen5,Y4260notseen6}, hidden
charm~\cite{Y4260notseen7,Y4260notseen8,Y4260notseen9} and charmless~\cite{Y4260notseen10,Y4260notseen11,Y4260notseen12} decay modes have be seen. Different interpretations were proposed to explain its structure, such as the  charmonium states
$4~^3{S_1}$~\cite{Y4260_4S_1,Y4260_4S_2,Y4260_4S_3} and $3~^3{D_1}$~\cite{Y4260_3D}, hybrid charmonium~\cite{Y4260_hybrid_1,Y4260_hybrid_2},
tetraquark state~\cite{Y4260_4q_1,Y4260_4q_2,Y4260_4q_3,Y4260_4q_4}, molecular
state~\cite{Y4260_m_1,Y4260_m_2,Y4260_m_3,Y4260_m_4,Y4260_m_5,Y4260_m_6}, and non-resonance explanation~\cite{Y4260_nonres1,Y4260_nonres2,Y4260_nonres3}.\\
\indent
The $Z_c(3900)$ was observed in the $J/\psi\pi^{\pm}$ invariant mass distribution of the $e^+e^-\to \pi^+\pi^-J/\psi$ process by the BESIII
Collaboration~\cite{PhysRevLett.110.252001}.
Subsequently, additional $Z_c(3900)$ decay channels were observed, including %$\pi^0 J/\psi$~\cite{PhysRevLett.115.112003},
$D\bar{D}^*+c.c$~\cite{PhysRevLett.112.022001,PhysRev.D92.092006,PhysRevLett.115.222002},
$D^*\bar{D}^*+c.c$~\cite{PhysRevLett.112.132001,PhysRevLett.115.182002}, and there are evidence for $Z_c(3900)$
decays into $\pi h_c$~\cite{PhysRevLett.111.242001,PhysRevLett.113.212002}.
The $J^P$ of $Z_c(3900)$ was determined to be $1^+$ with
a partial wave analysis of the $\pi^+\pi^-J/\psi$ final state~\cite{Collaboration:2017njt}.
Since its discovery, many interpretations on the nature of the $Z_c(3900)$
have been proposed, such as a $D\bar{D}^*$ molecule~\cite{ddmolecule}, a tetra-quark state~\cite{tetraquark}, a cusp effect~\cite{cusps}, and dynamical generation through threshold effects~\cite{threshold1,threshold2}.

Despite many interpretations, the natures of the $Y(4260)$ and $Z_c(3900)$ are still unclear. To comprehend these states, it is necessary to study more decay  modes. All of the observed decay modes of the $Y(4260)$ and $Z_c(3900)$ are associated with the charm sector and no
light hadron decay modes have been found
yet~\cite{lhd1,lhd2,PhysRev.D92.032009}. A search for light hadron decay modes of the $Y(4260)$ and $Z_c(3900)$ is complementary to previous studies and may help to
distinguish between different theoretical models and to understand strong interaction effects in this energy region.
 Among the large number of potential light hadron decay modes, the branching fractions~(BFs) of charmonium states decaying into $K_S K^\pm
\pi^\mp \pi^0$ and $K_S K^\pm \pi^\mp \eta$ are usually large~\cite{PDG}.
In four-body final state, there should be abundant intermediate
states, which may supply more possible decay channels for searching $Y(4260)$ and $Z_c(3900)$.
Furthermore, the existence of charged and
neutral pions in the final states enable a study of isospin multiplets. In this paper, we present a
measurement of the Born
cross sections~($\sigma_{\rm{B}}$) for $e^+e^-\to K_S^0K^{\pm}\pi^\pm\pi^0$ and
$K_{S}^{0}K^{\pm}\pi^{\mp}\eta$.
We also report upper limits
of $e^+e^-$ $\to$ $Y(4260)\to K_S^0K^\pm\pi^\mp\pi^0$, $e^+e^-$ $\to$ $Y(4260)\to
K_S^0K^\pm\pi^\mp\eta$, $e^+e^-\to \pi^{\mp,0}Z_c(3900)^{\pm,0}\to K_S^0K^\pm\pi^{\mp} \pi^{0}$,
and $e^+e^-\to \pi^{\mp}Z_c(3900)^{\pm} \to K_S^0K^\pm \pi^{\mp}\eta$.

\section{\texorpdfstring{Detectors and Data samples}{Detector and MC Simulation}}

The BESIII detector~\cite{Ablikim2010345}, operating at the BEPCII collider~\cite{BEPCII}, is a general purpose
spectrometer with a geometrical acceptance of 93\% of 4$\pi$ solid angle. It has four main
components: (1) a small-cell, helium-based (60\% He, 40\% $\mbox{C}_3\mbox{H}_8$) multi-layer drift
chamber (MDC) with 43 layers providing an average single-hit resolution of
135\,$\mu$m, a momentum resolution of $0.5\%$ at 1.0\,GeV/$c$ in a 1.0\,T magnetic field, and a specific ionization energy loss ($dE/dx$) resolution better than 6\%, (2) a time-of-flight
(TOF) detector constructed of 5 cm thick plastic scintillators, with 176 strips of
2.4\,m length in two layers in the barrel and 96 fans of the
end-caps with time resolutions of 80 and 110\,ps,
respectively, which provide a
2$\sigma$ $K/\pi$ separation for momenta up to $\sim$ 1.0\,GeV/$c$, (3) an electromagnetic
calorimeter (EMC) consisting of 6240 CsI (Tl) crystals in a cylindrical barrel structure
and two end-caps with an energy resolution of 2.5\% (5\%) at 1.0\,GeV and a position
resolution of 6\,mm (9\,mm) in the barrel (end-caps), and (4) a muon counter (MUC) consisting
of resistive plate chambers (RPCs) in nine barrel and
eight end-cap layers, which provide a 2\,cm position resolution. More details of the BESIII detector can be found in Ref.~\cite{Ablikim2010345}.

This analysis is based on $5.2 \ \mathrm{fb}^{-1}$ $e^+ e^-$ annihilation data samples~\cite{Ablikim:2015nan} collected with the BESIII detector at center-of-mass energies ($\sqrt{s}$) from
3.90 to 4.60 GeV~\cite{Ablikim:2015zaa}, which are listed in
Table~\ref{tab:result_pi0}.
Monte-Carlo (MC) simulations are used to optimize the event selection criteria, to study the detector acceptance and to understand the potential backgrounds. The {\sc
geant4}-based~\cite{geant4} simulation software {\sc boost}~\cite{ref:boost} is implemented to simulate the detector response, describe geometry and material,
realize
digitization, and incorporate time-dependent beam backgrounds. Six generic MC samples, equivalent to the integrated luminosity of the data at the energy points $4.009$, $4.230$, $4.260$, $4.360$, $4.420$ and $4.600$ $\mathrm{GeV}$ are generated to study the backgrounds.
The primary known decay channels are generated using {\sc evtgen}~\cite{EVTGEN2} with the BFs set to the world average values~\cite{PDG}
while the unknown decay modes are generated with {\sc lundcharm}~\cite{EVTGEN}. Continuum hadronic events are generated with
{\sc kkmc}~\cite{KKMC} and
QED processes such as Bhabha scattering, dimuon, and digamma events are generated with {\sc kkmc} and {\sc babayaga}~\cite{Babayaga}. To
study the efficiency of each final state, a sample of $1\times10^5$ signal events is generated at each energy point using {\sc kkmc}, which
simulates $e^+ e^-$ annihilation, including beam energy spread and ISR effects.

%===========================================================================
%=========================== Data analysis=================
%===========================================================================

\section{Data analysis}

\subsection{{\boldmath Measurement of $\sigma_{\rm{B}}(e^+e^- \to K_{S}^{0}K^{\pm}\pi^{\mp}\pi^0)$ and $\sigma_{\rm{B}}(e^+e^- \to K_{S}^{0}K^{\pm}\pi^{\mp}\eta)$}}
\label{section:A}
Candidate events for $e^+e^- \to K_{S}^{0}K^{\pm}\pi^{\mp}\pi^0/\eta$, with $K_S^0 \to \pi^+ \pi^-$ and $\pi^0/\eta \to \gamma \gamma$
are selected according to the following steps.
First, $K_{S}^{0}$ candidates
are selected by looping over all pairs of oppositely charged tracks, which are assumed to be pions.
Next, primary and secondary vertex
fits~\cite{SecondVFit} are performed and the decay length of the secondary vertex fit is required to be greater than twice its uncertainty.
Furthermore, the invariant mass of the pion pair is required to satisfy $|M(\pi^{+}\pi^{-})-M_{K_{S}^{0}}|< 12$~MeV/$c^{2}$, where $M_{K_{S}^{0}}$ denotes the
nominal mass of the $K_{S}^{0}$~\cite{PDG}. If there are multiple $K_{S}^{0}$ candidates in one event, the one with the smallest $\chi^{2}$ from the secondary vertex
fit is selected.

In addition to the two charged tracks that make up the $K_S^0$, two oppositely charged tracks are required. For the latter two charged tracks, the polar angle $\theta$ must satisfy $|\cos\theta| < 0.93$ and the distance of closest approach to the interaction point must be less than $10.0 ~\mathrm{cm}$ and $1.0~\mathrm{cm}$ along the
beam direction and in the plane
perpendicular to the beam direction, respectively. The particle type for each charged track is determined by selecting the hypotheses with the highest probability, which
is calculated with the
combined information from TOF and $dE/dx$ measurements for different particle hypotheses. One charged track must be identified as a kaon and the other as a pion.

Photons are reconstructed from clusters deposited in the EMC, with the energy measured in the TOF included to improve reconstruction
efficiency
and energy resolution. At least two photons are required per event. The energy of a photon candidate is required to be larger than $25 ~\mathrm{MeV}$ in the barrel region ($|\cos\theta| < 0.80$) or $50
~\mathrm{MeV}$ in the end-cap region ($0.86 < |\cos\theta| < 0.92$). The cluster timing is required to be between $0$ and $700~\mathrm{ns}$ to suppress
electronic noise and energy depositions unrelated to the event of interest. To eliminate showers associated with charged particles, the opening angle between a photon candidate
and the extrapolated position of the closest charged track should be larger than $20$ degrees.

Finally, a four-constraint ($4\mathrm{C}$) kinematic fit imposing energy-momentum conservation is performed to the final states. Only events with
$\chi^2_{\rm{4C}}<60$ are accepted. For
events with more than two photon candidates, the photon pair with the smallest $\chi^2_{4\mathrm{C}}$ from the kinematic fit is accepted.
\begin{figure}[htbp]
\subfigure{
    \includegraphics[width = 0.4\textwidth]{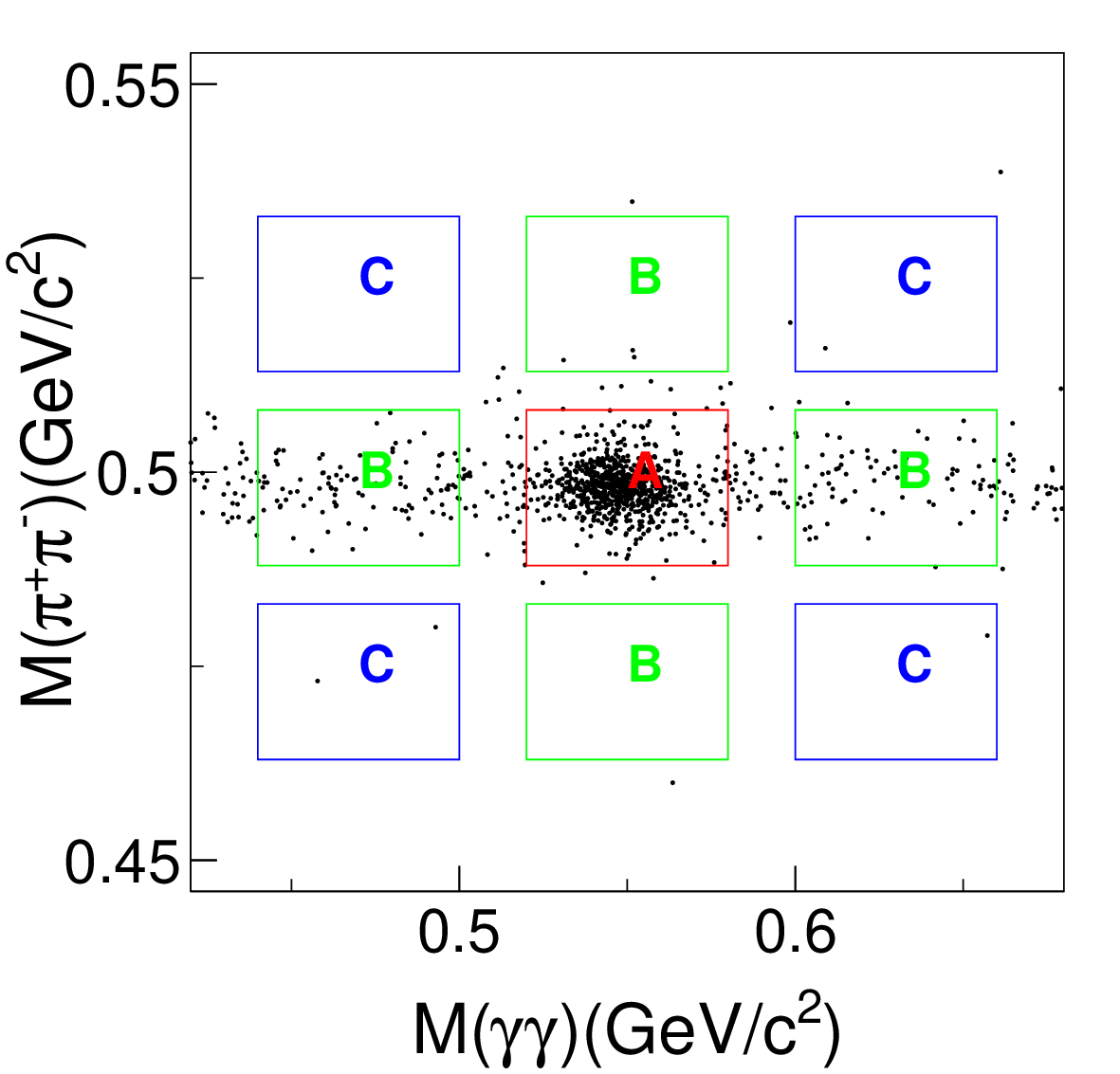}\put(-100,170){\bf(a) }
    }
\subfigure{
    \includegraphics[width = 0.4\textwidth]{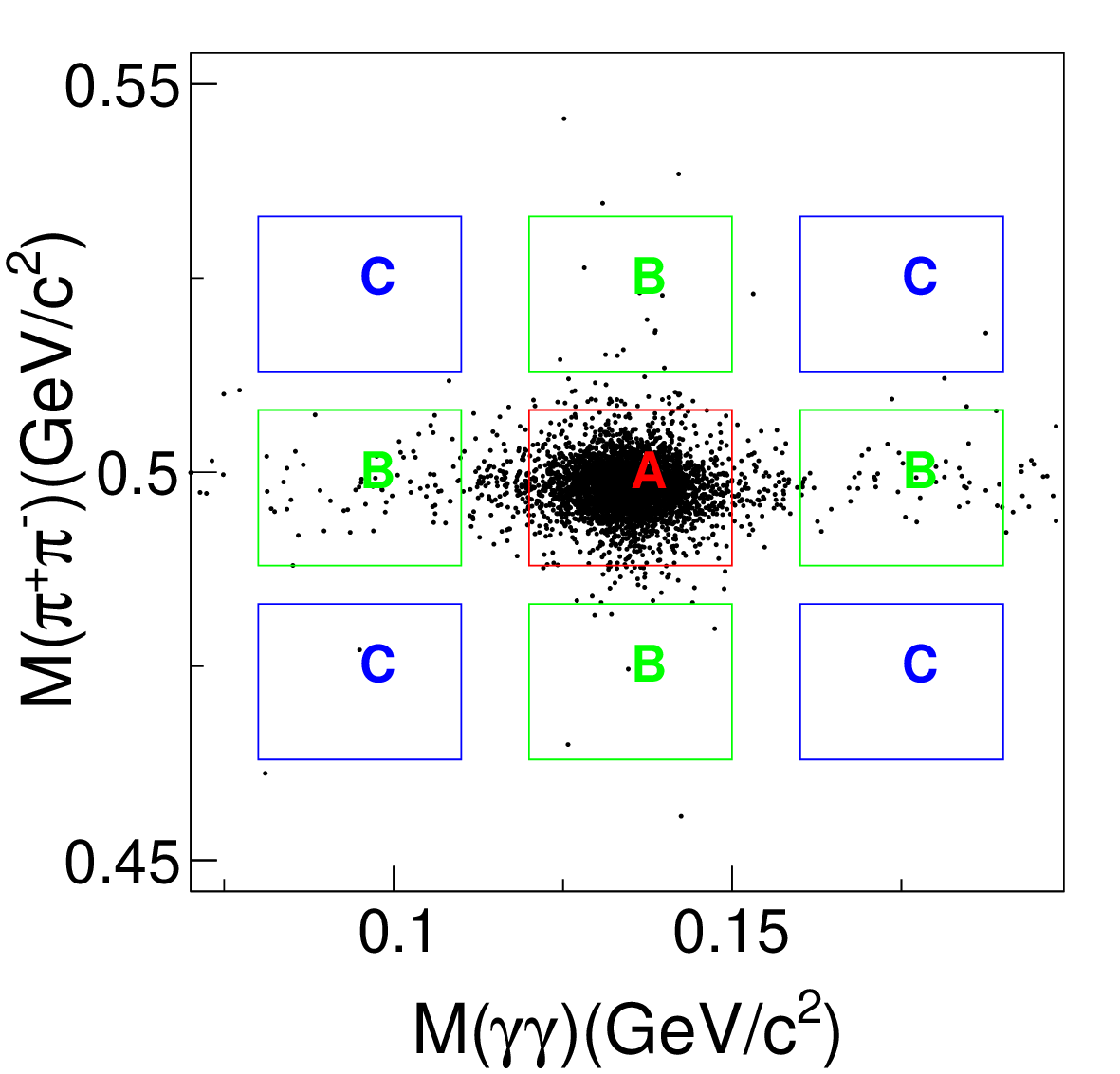}\put(-100,170){\bf(b) }
    }
\caption{(Color online) The distributions of M($\pi^+ \pi^-$) versus M($\gamma \gamma$) at $\sqrt{s}=4.258$ GeV. The top is the $\pi^0$ mode and the bottom is the $\eta$ mode. The boxes with mark ``A'' is the signal region and the boxes with mark ``B'' and ``C'' are sideband regions.}\label{fig1}
\end{figure}
After the 4C kinematic fit, no peaking background is observed in the generic MC samples. The invariant mass distributions of $\pi^+ \pi^-$ versus
$\gamma \gamma$ at $4.258~\mathrm{GeV}$ are shown in Fig.~\ref{fig1} as an example in which obvious $K_S^0$ and $\pi^0/\eta$ peaks are observed. The signal regions are
defined as $M(\pi^+\pi^-)\in (0.488, 0.508) ~\mathrm{GeV}/c^2$, $M(\gamma \gamma) \in (0.12, 0.15)~\mathrm{GeV}/c^2$ (for the $\pi^0$ mode) and $M(\gamma \gamma) \in
(0.52, 0.58)~\mathrm{GeV}/c^2$ (for the $\eta$ mode). The sideband regions are defined as $M(\pi^+\pi^-) \in (0.463, 0.483) \cup (0.513, 0.533) ~\mathrm{GeV}/c^2$,
$M(\gamma \gamma) \in (0.08, 0.11)\cup(0.16, 0.19)~\mathrm{GeV}/c^2$ (for the $\pi^0$ mode) and $M(\gamma \gamma) \in (0.44, 0.50)\cup(0.60, 0.66)~\mathrm{GeV}/c^2$
(for the $\eta$ mode). The signal yields at each energy, presented in
Tables~\ref{tab:result_pi0} and \ref{tab:result_eta}, are obtained according to $N_{\rm{sig}} = N_{\rm{A}}- \sum N_{\rm{B}}/2 + \sum N_{\rm{C}}/4$, where $N$ is the
number of events and the subscript A denotes the
signal region, and the subscripts B and C denote the sideband regions.

The Born cross section is calculated from
\begin{equation}
\sigma_{\rm{B}} = \frac{N_{\mathrm{sig}}}{\mathcal{L}\cdot\mathcal{B}\cdot\epsilon\cdot(1+\delta^{\mathrm{ISR}})\cdot\frac{1}{|1-\Pi(s)|^2}}\;, \label{eq:sigma}
\end{equation}
where $\mathcal{L}$ is the integrated luminosity, $\epsilon$ is the detection efficiency, $\mathcal{B}$ is the product of the BF of $K_S^0 \to \pi^+ \pi^-$
and that of $\pi^0/\eta \to \gamma \gamma$
~\cite{PDG}, $\frac{1}{|1-\Pi(s)|^2}$ is
the vacuum polarization correction factor~\cite{VP}, and $(1+\delta^{\rm{ISR}})$ is the ISR correction factor~\cite{Kuraev:1985hb}
which is determined by the MC simulation programmer {\sc kkmc}. The ISR factors are set to 1.0 to get the initial cross section
lineshape as input to {\sc kkmc}. From {\sc kkmc}, the updated ISR factors are obtained, then the cross section lineshape is updated too.
We repeat this process till both ISR factors and cross section converge.

\begin{figure}[htbp]
\subfigure{
    \includegraphics[scale=0.2]{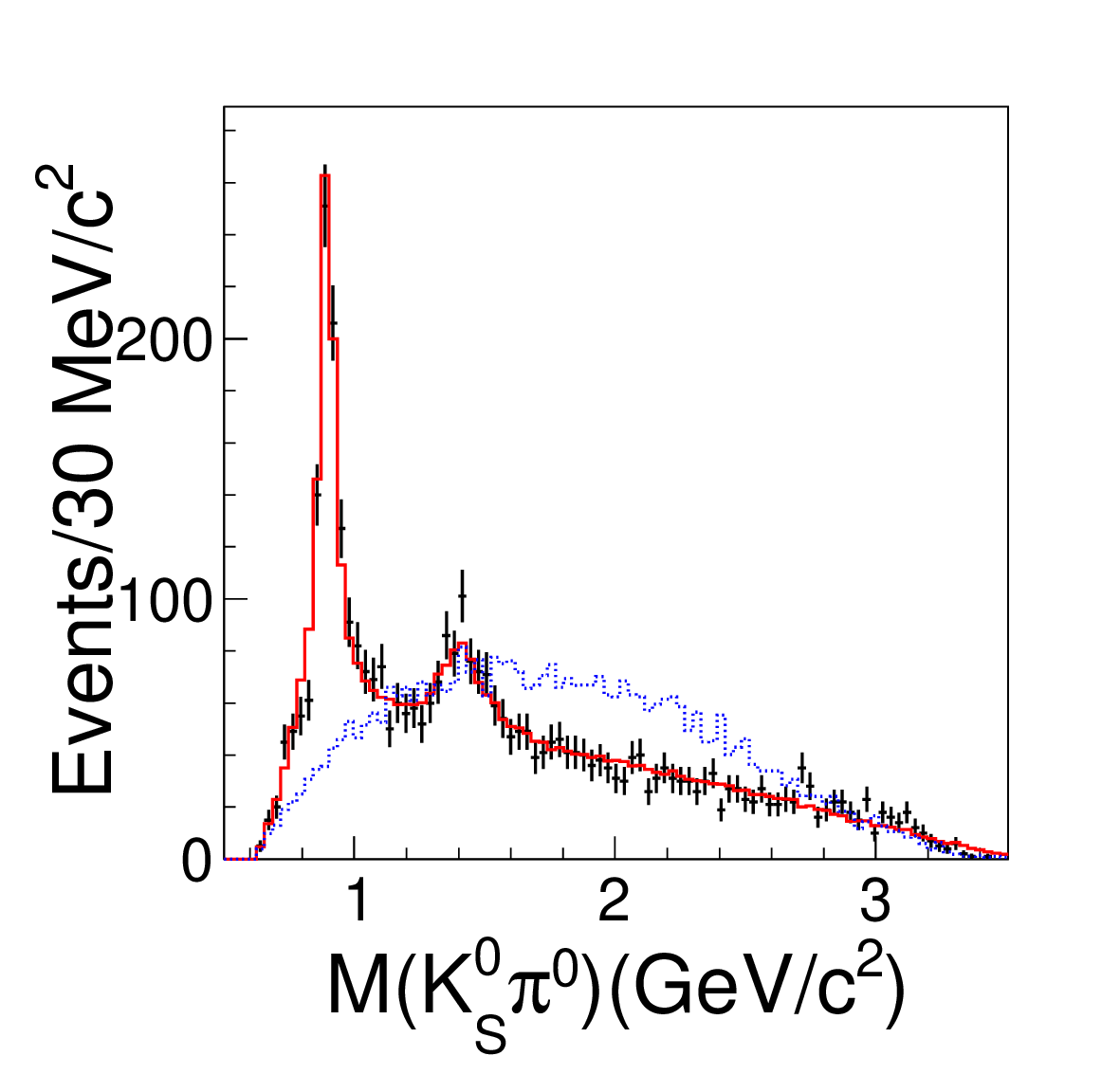}\put(-55,85){\bf(a) }
    }
\subfigure{
    \includegraphics[scale=0.2]{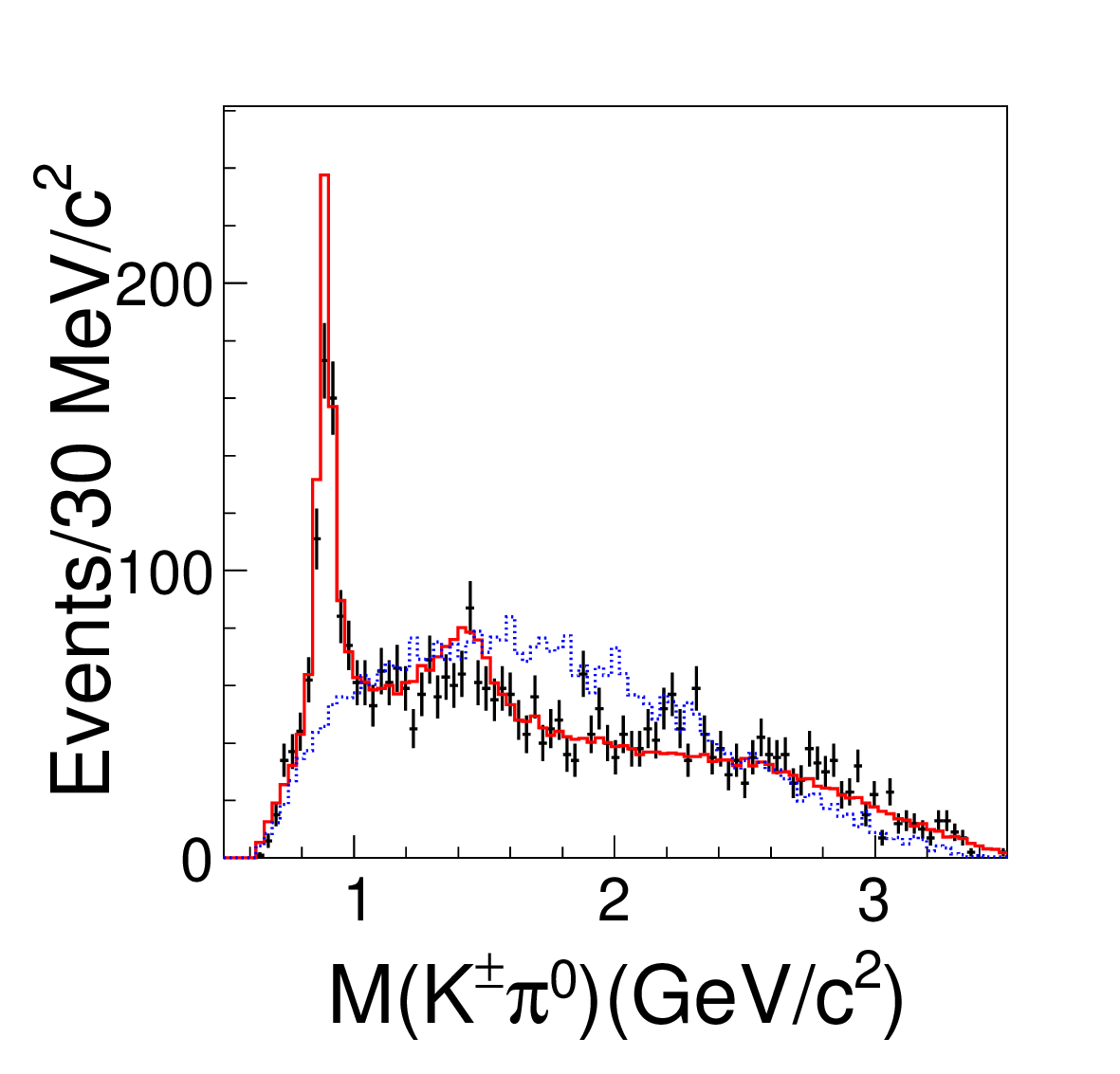}\put(-55,85){\bf(b) }
    }
\subfigure{
    \includegraphics[scale=0.2]{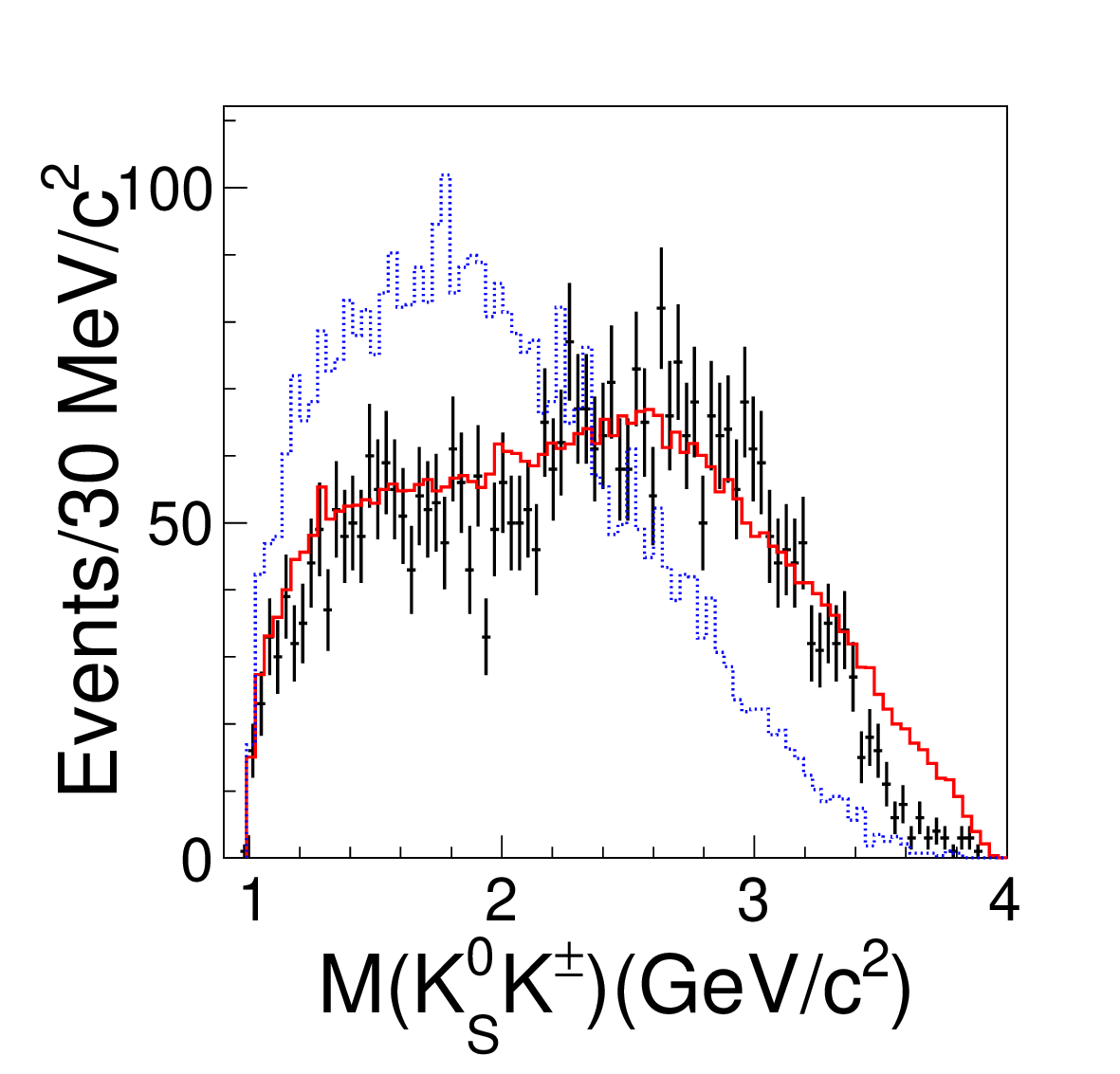}\put(-55,85){\bf(c) }
    }
\subfigure{
    \includegraphics[scale=0.2]{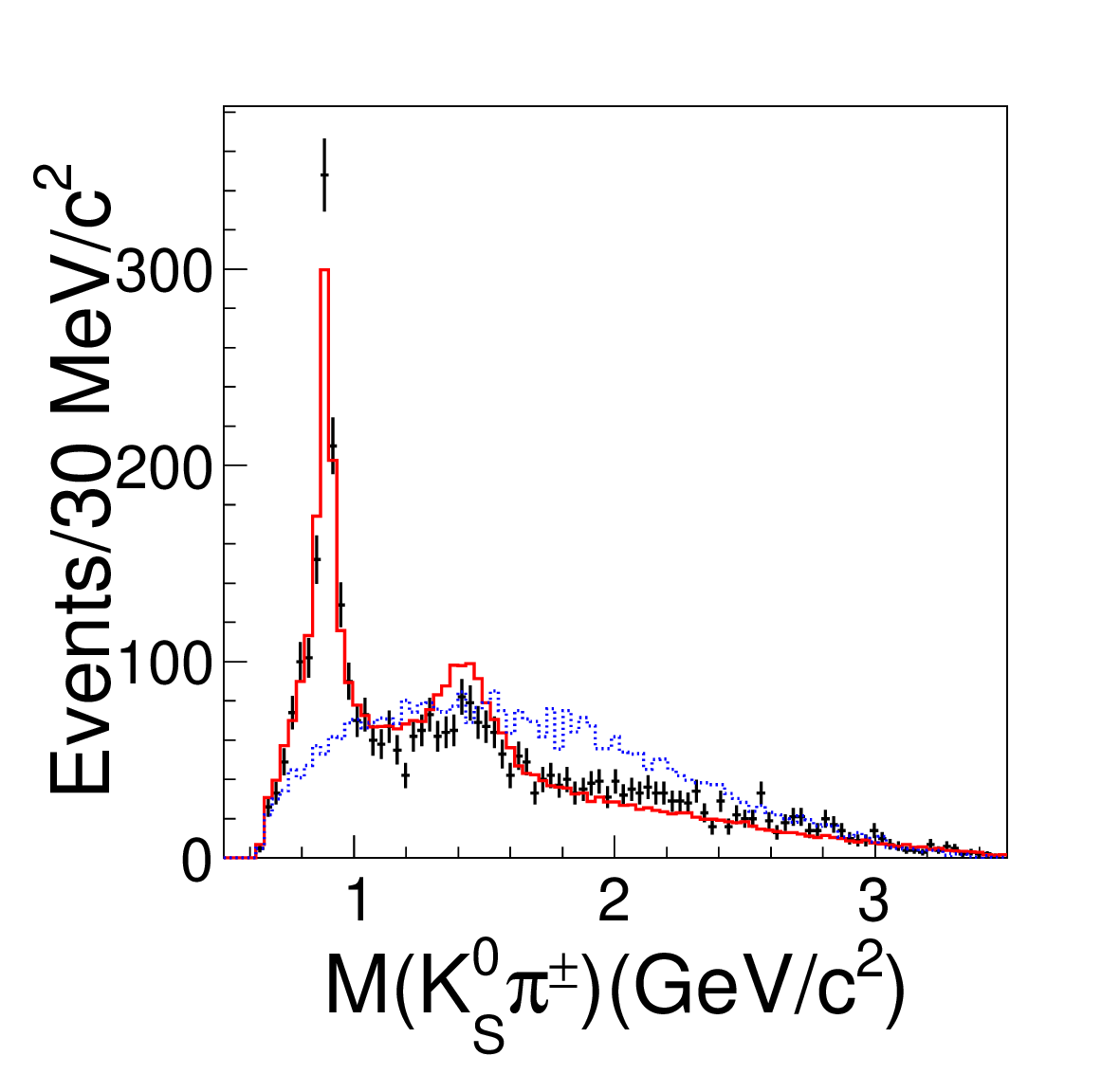}\put(-55,85){\bf(d) }
    }
\subfigure{
    \includegraphics[scale=0.2]{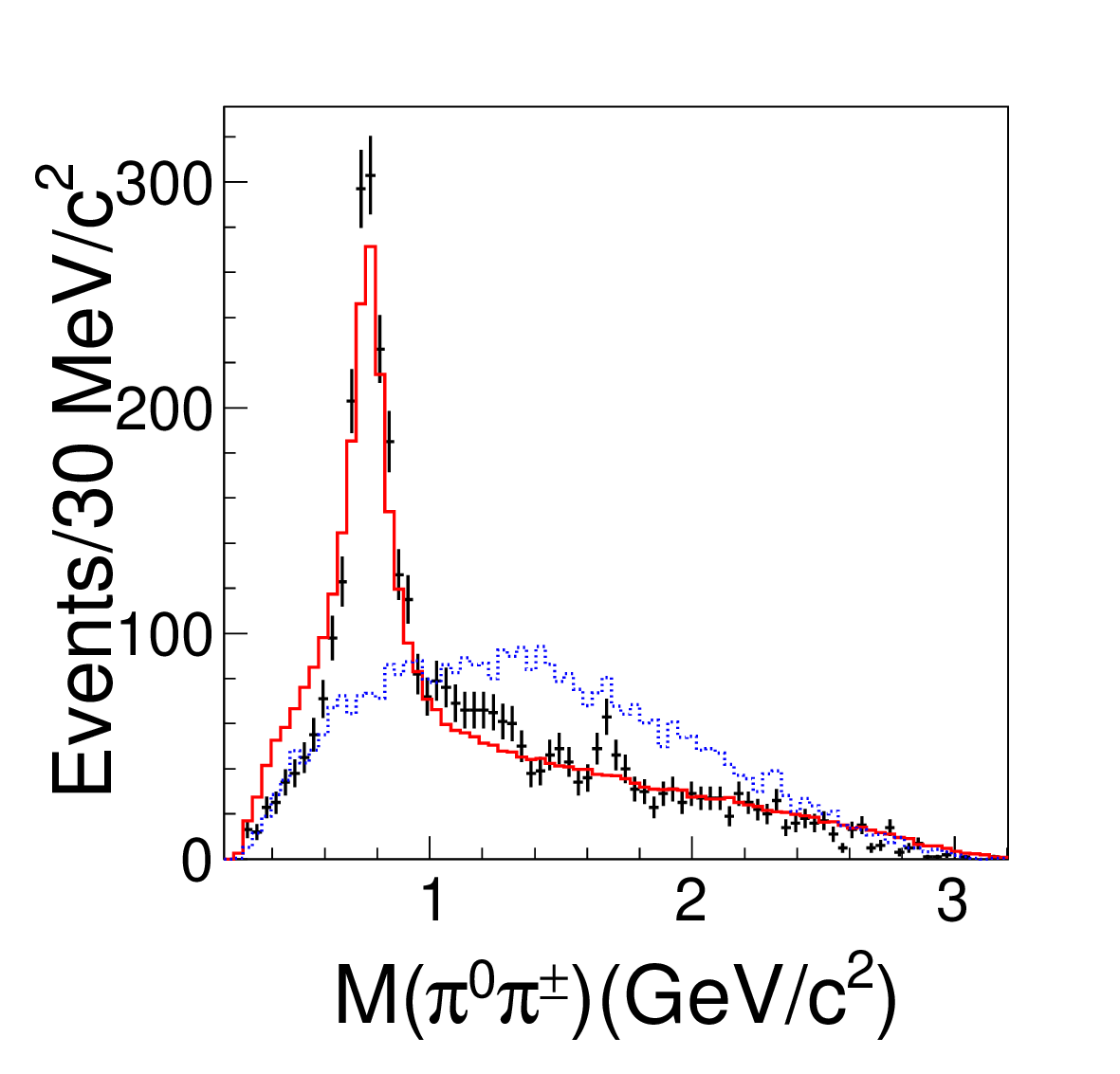}\put(-55,85){\bf(e) }
    }
\subfigure{
    \includegraphics[scale=0.2]{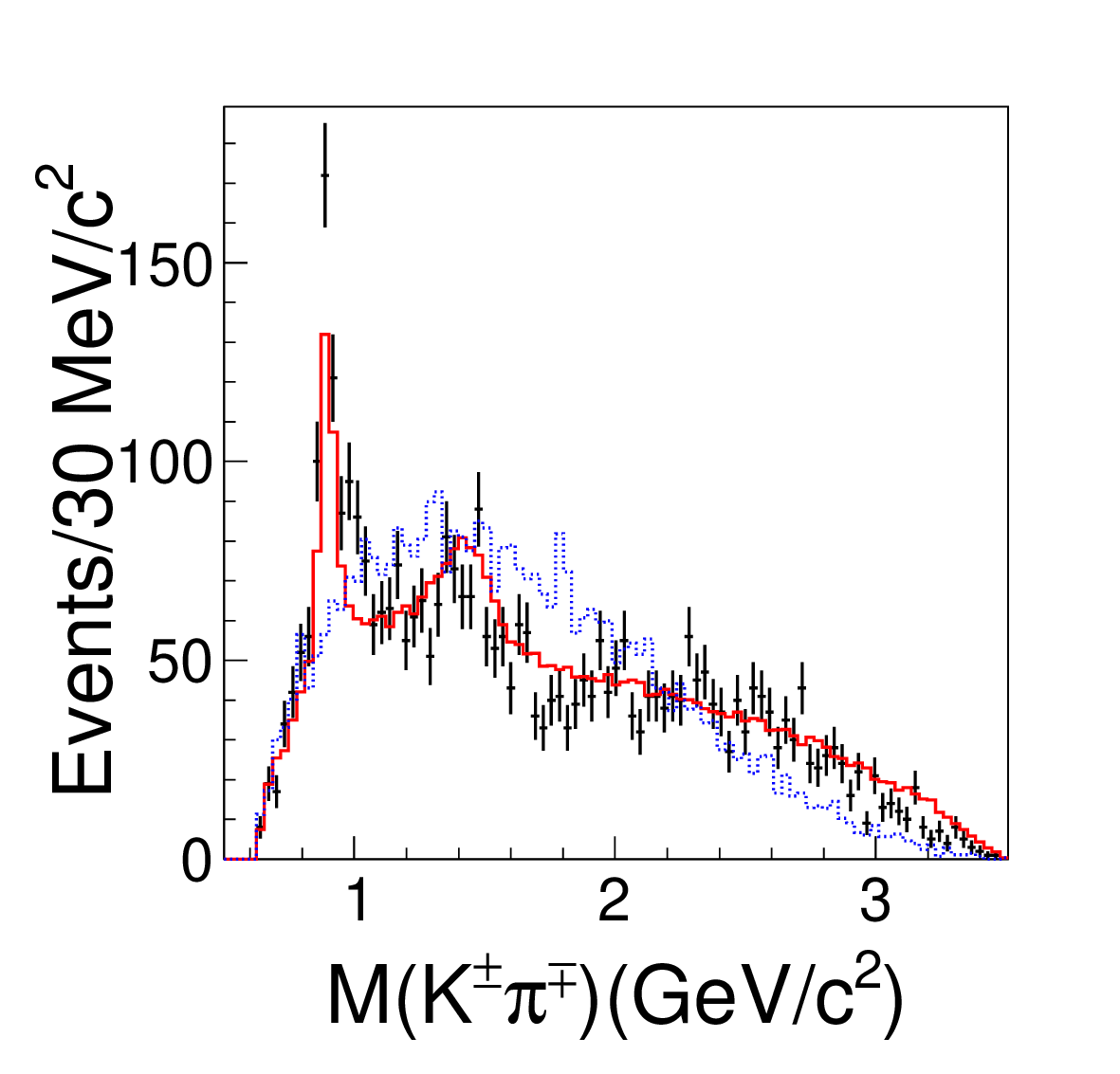}\put(-55,85){\bf(f) }
    }
\subfigure{
    \includegraphics[scale=0.2]{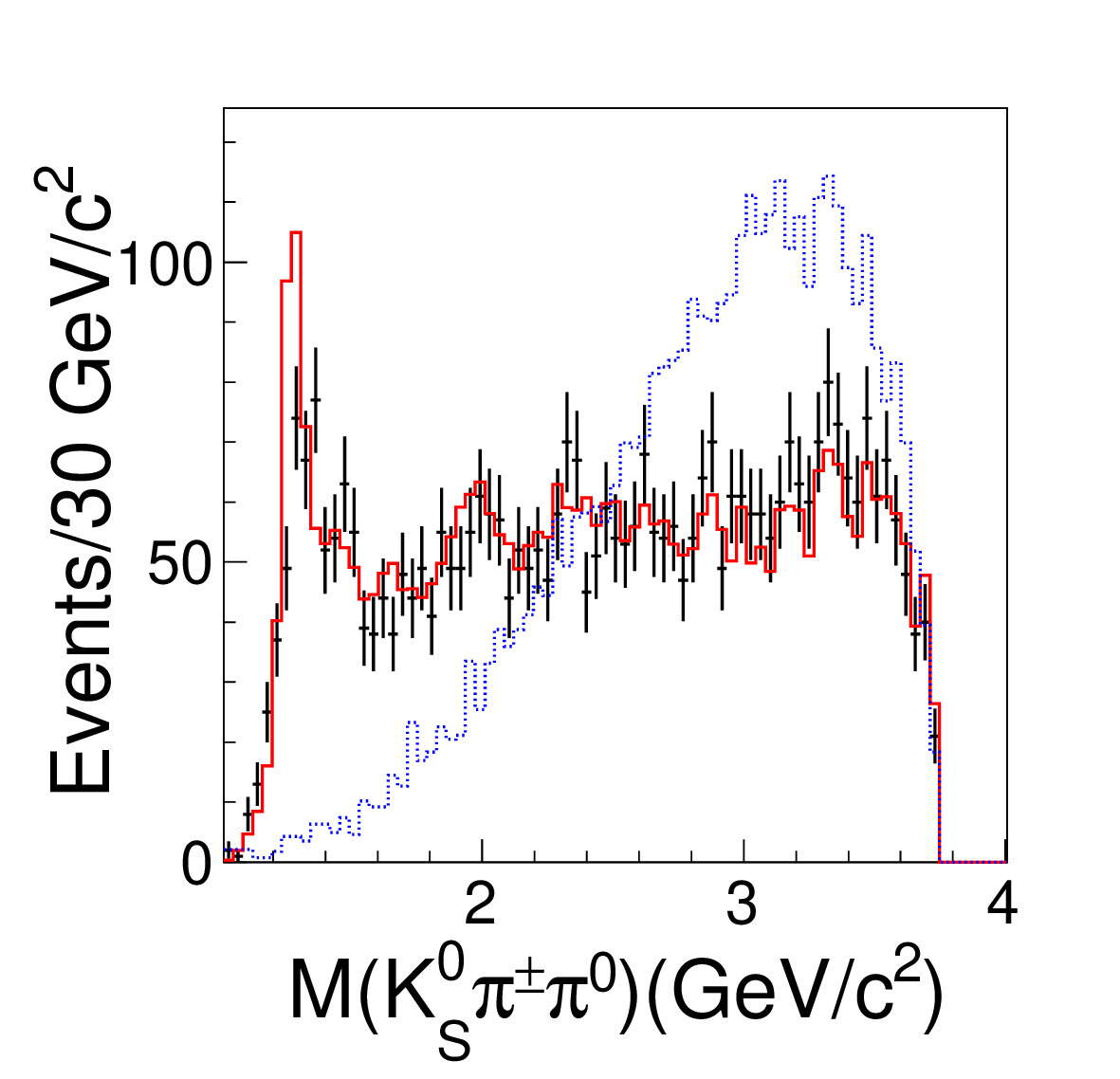}\put(-55,85){\bf(g) }
    }
\subfigure{
    \includegraphics[scale=0.2]{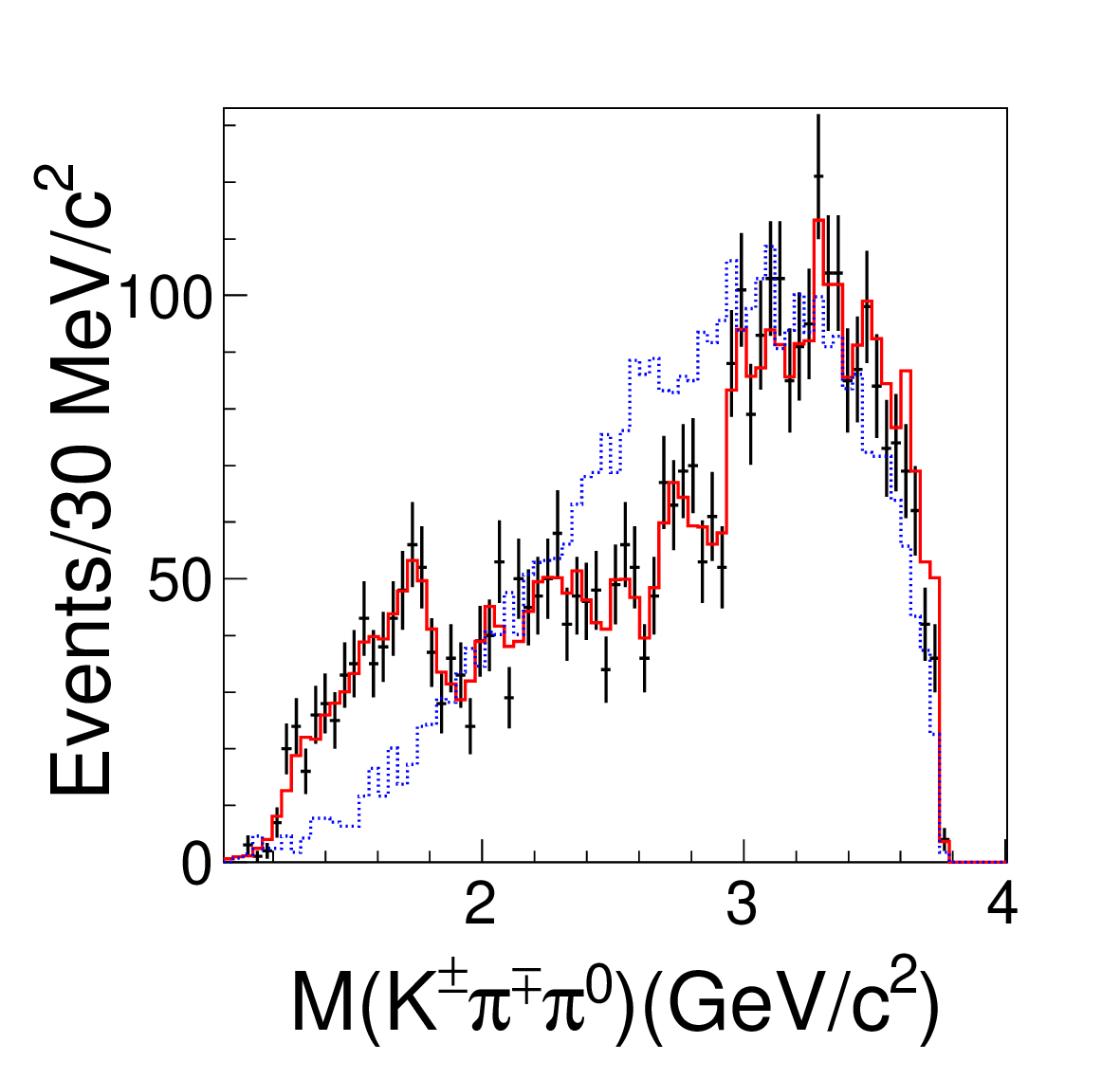}\put(-55,85){\bf(h) }
    }
\subfigure{
    \includegraphics[scale=0.2]{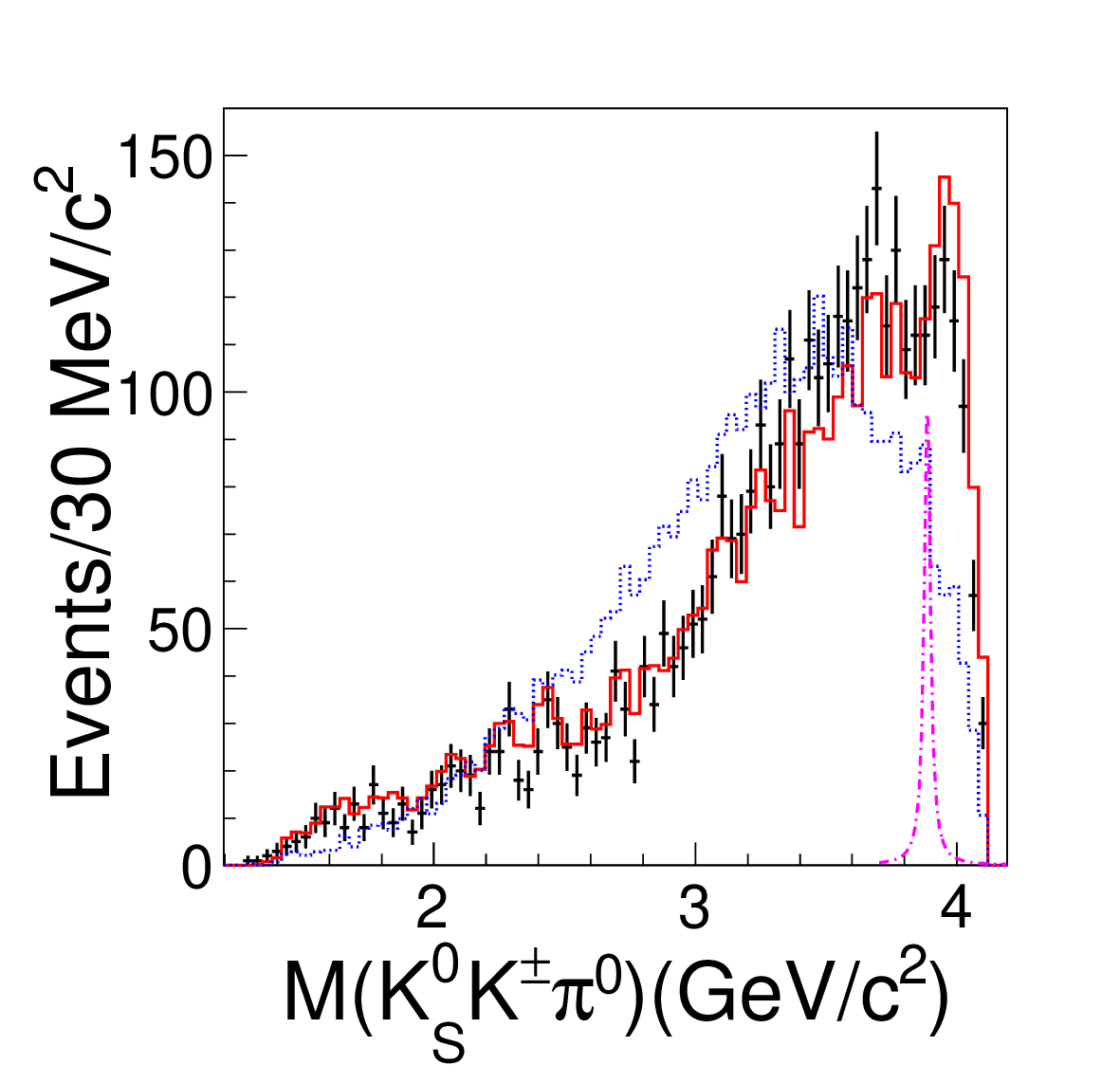}\put(-55,85){\bf(i) }
    }
\subfigure{
    \includegraphics[scale=0.2]{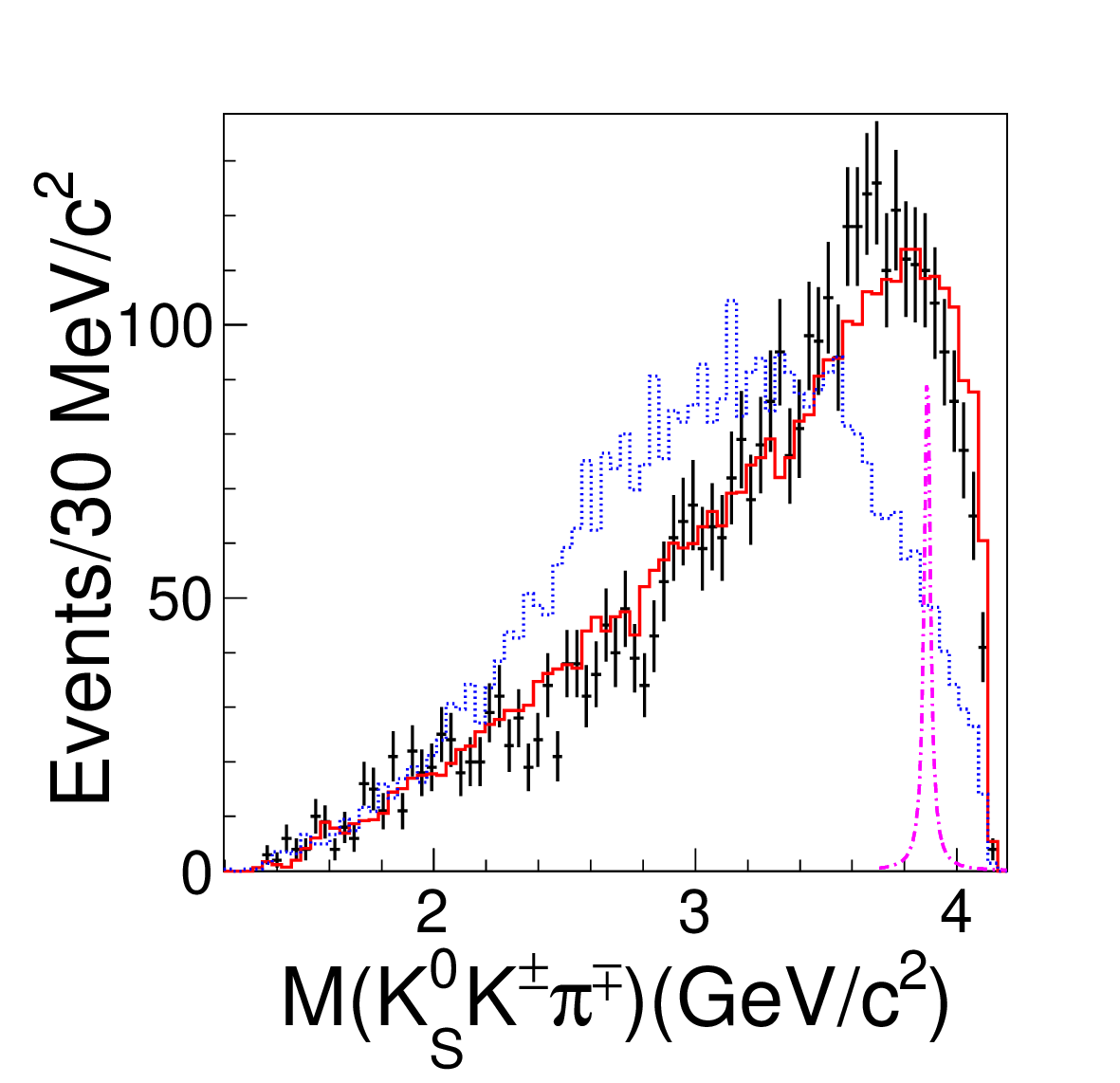}\put(-55,85){\bf(j) }
    }
\caption{(Color online) Invariant mass distributions of any two or three final state particles for the $\pi^0$ mode at 4.258 GeV. The black dots with error bars are the data. The red solid lines are the mixing MC sample. The blue dashed lines are the PHSP MC sample. The pink dash-dotted lines in plot~(i) and plot~(j) are the MC shape of the $Z_c(3900)$ with an arbitrary scale. \label{fig:resonances1}}
\end{figure}

\begin{figure}[htbp]
\subfigure{
    \includegraphics[scale=0.2]{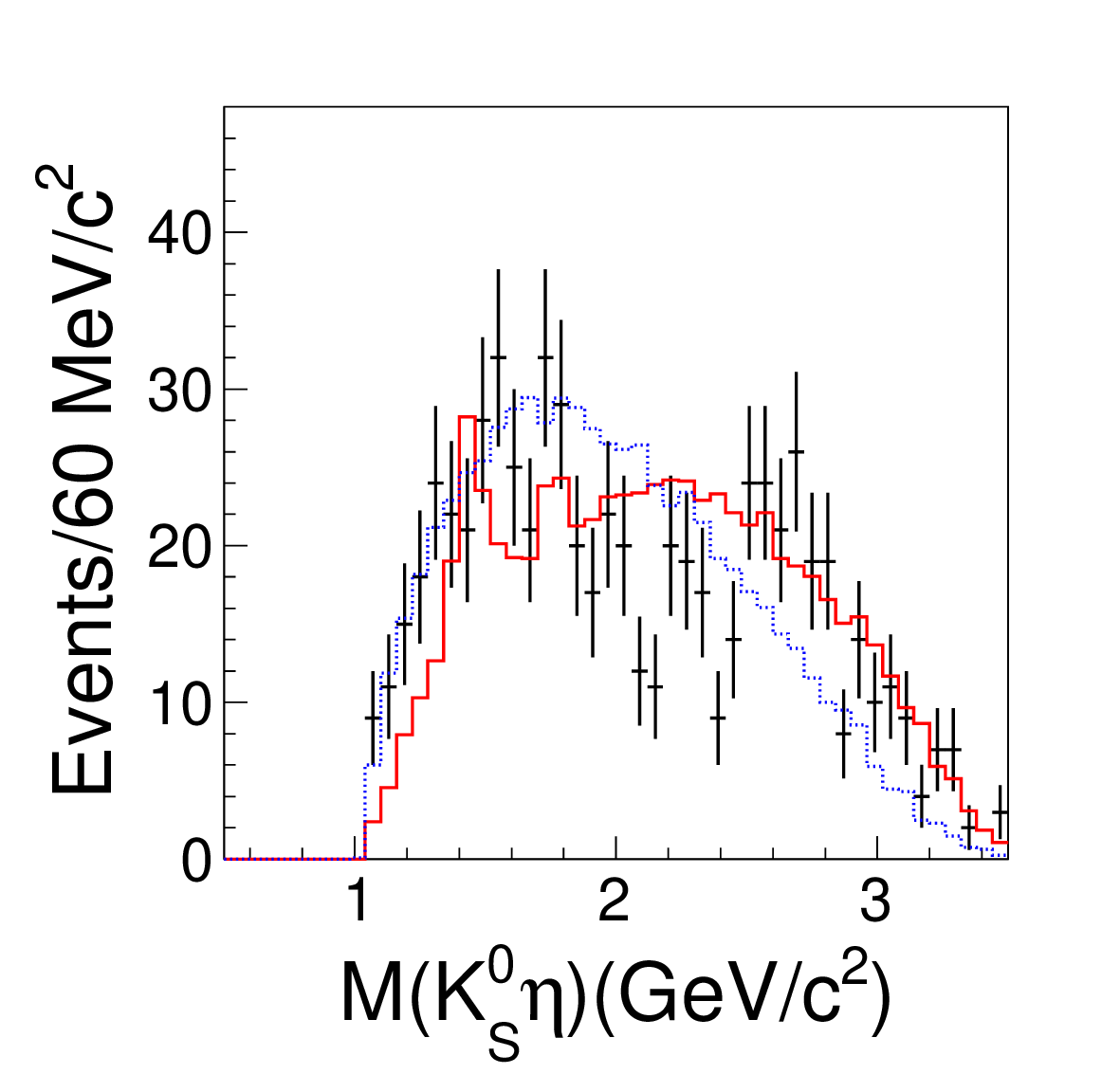}\put(-55,85){\bf(a) }
    }
\subfigure{
    \includegraphics[scale=0.2]{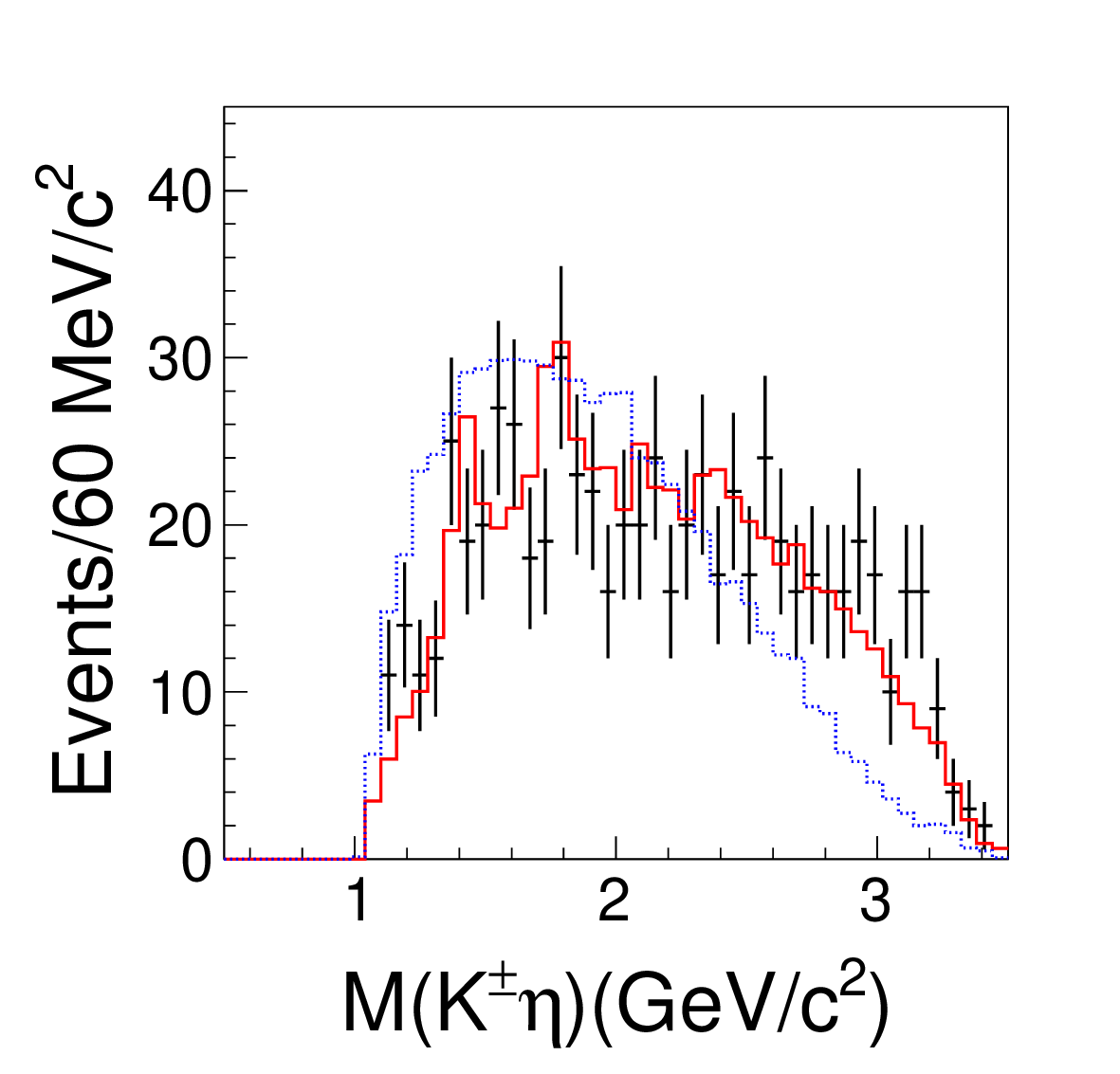}\put(-55,85){\bf(b) }
    }
\subfigure{
    \includegraphics[scale=0.2]{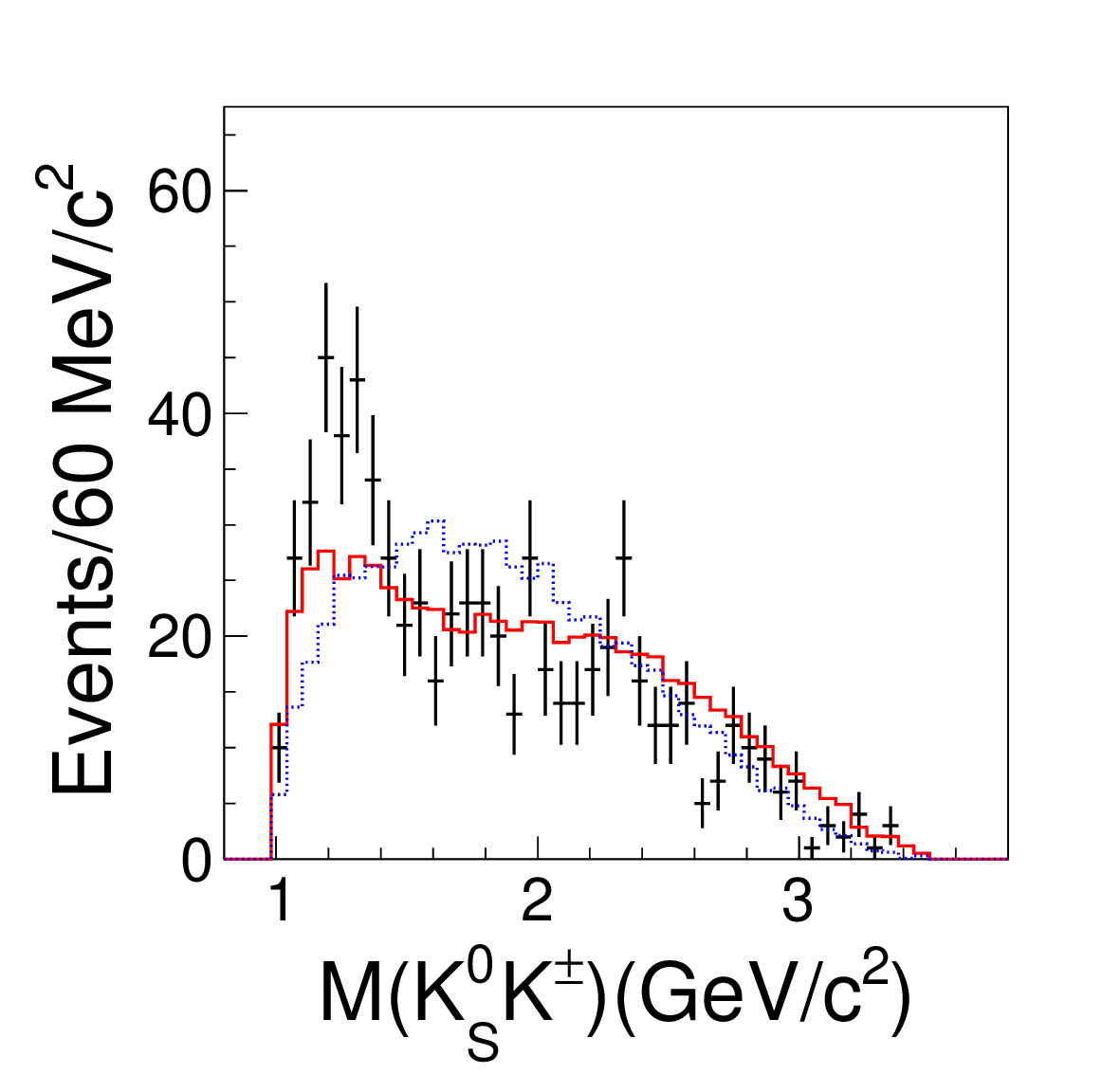}\put(-55,85){\bf(c) }
    }
\subfigure{
    \includegraphics[scale=0.2]{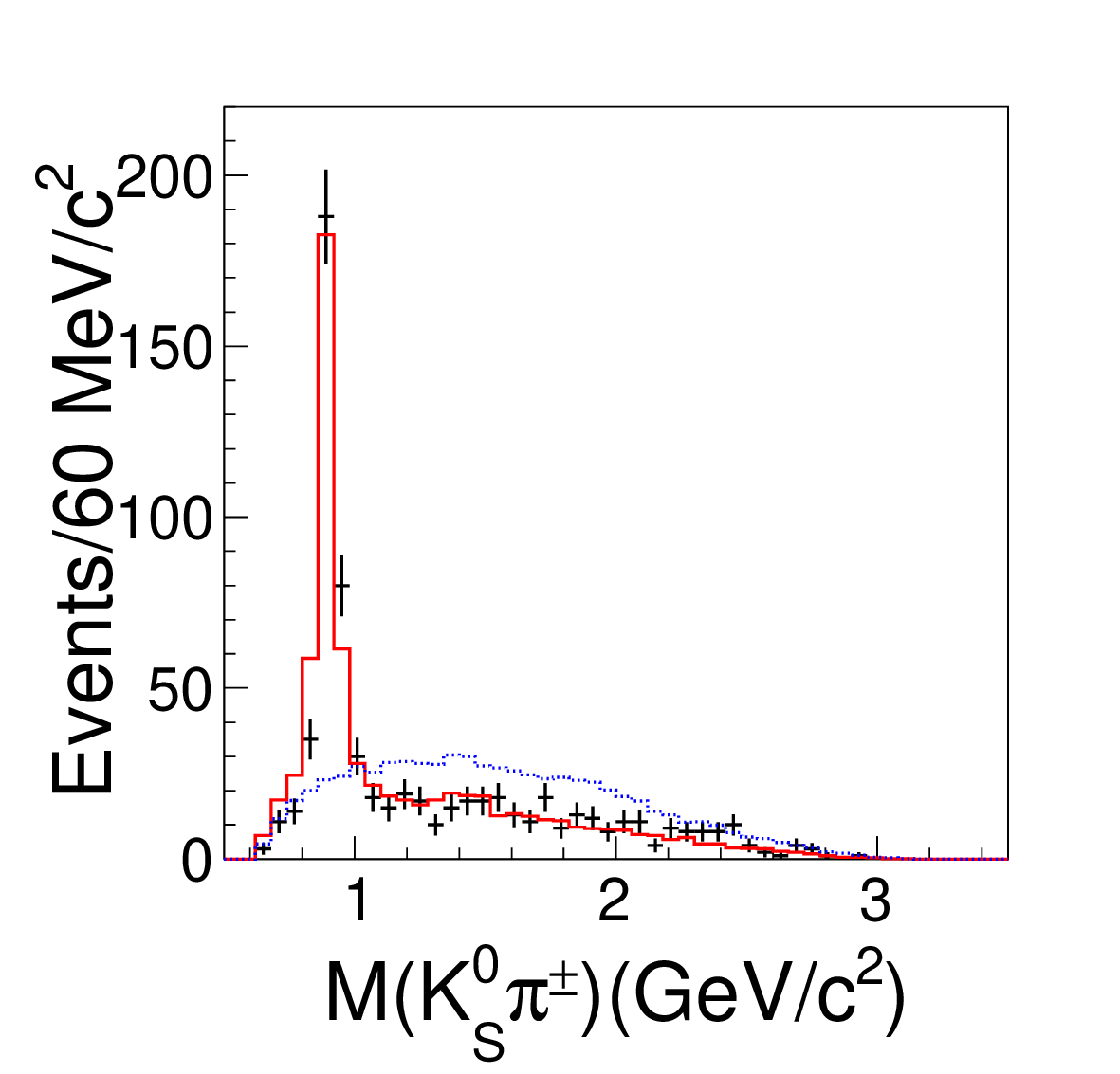}\put(-55,85){\bf(d) }
    }
\subfigure{
    \includegraphics[scale=0.2]{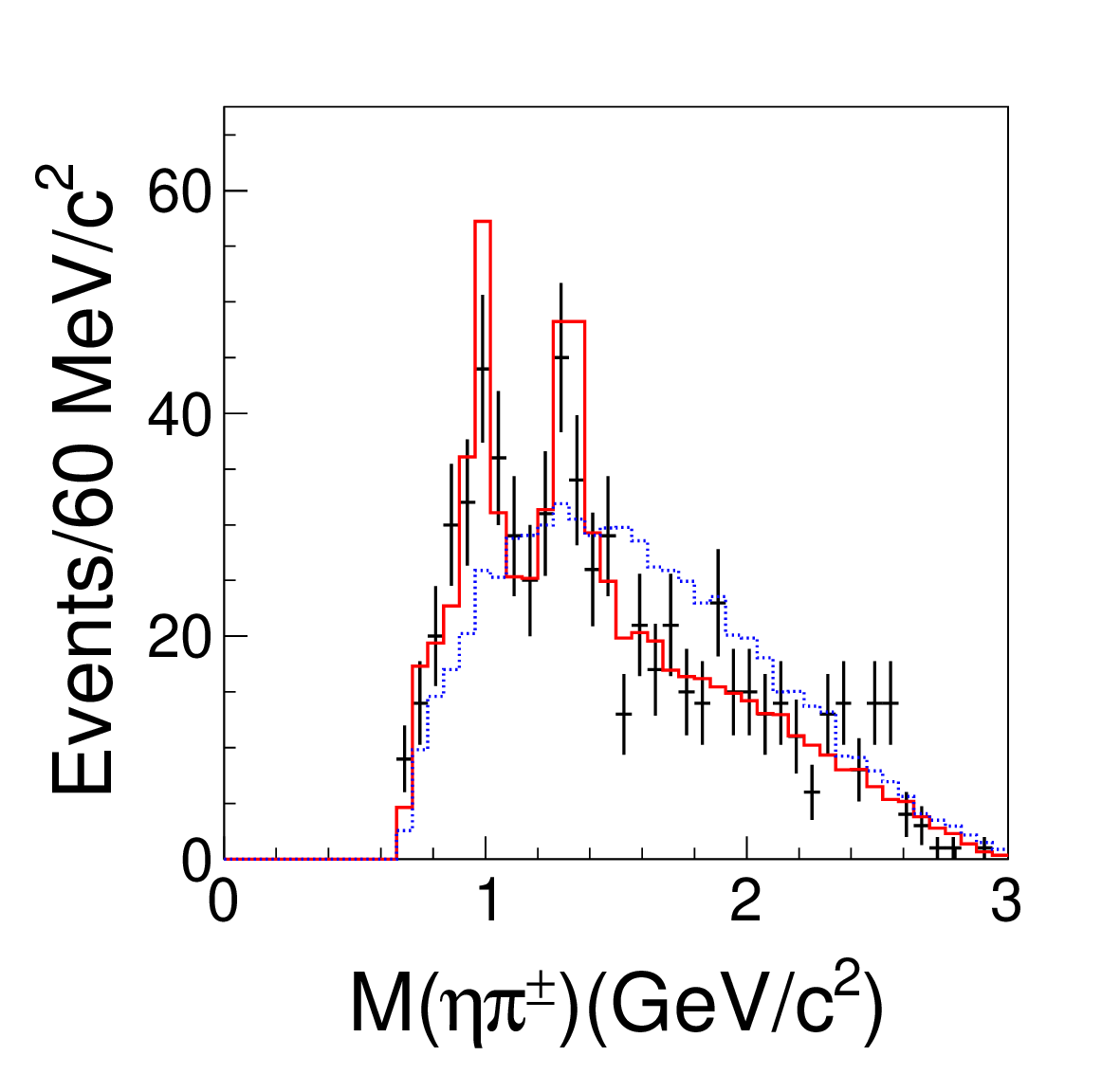}\put(-55,85){\bf(e) }
    }
\subfigure{
    \includegraphics[scale=0.2]{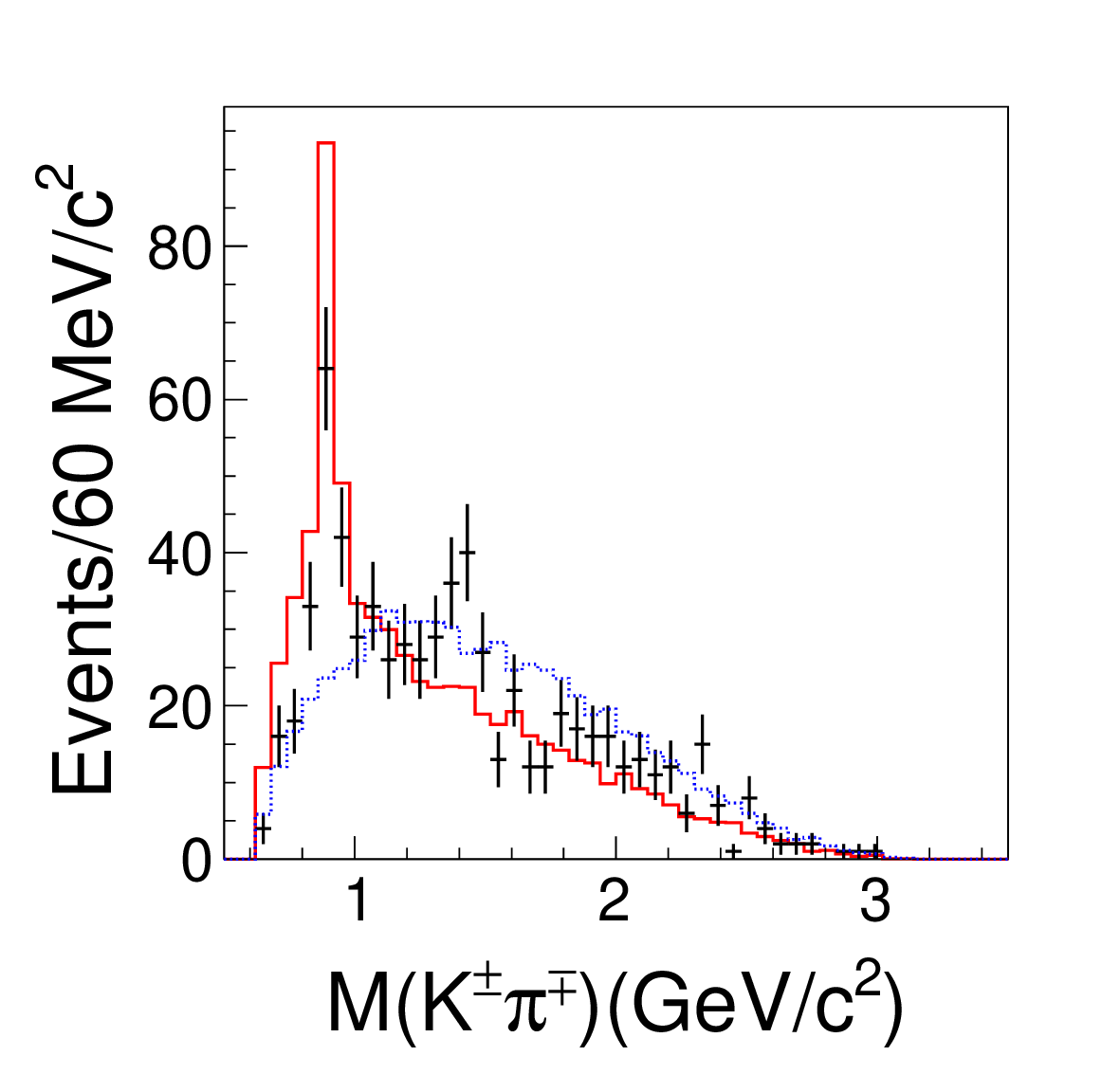}\put(-55,85){\bf(f) }
    }
\subfigure{
    \includegraphics[scale=0.2]{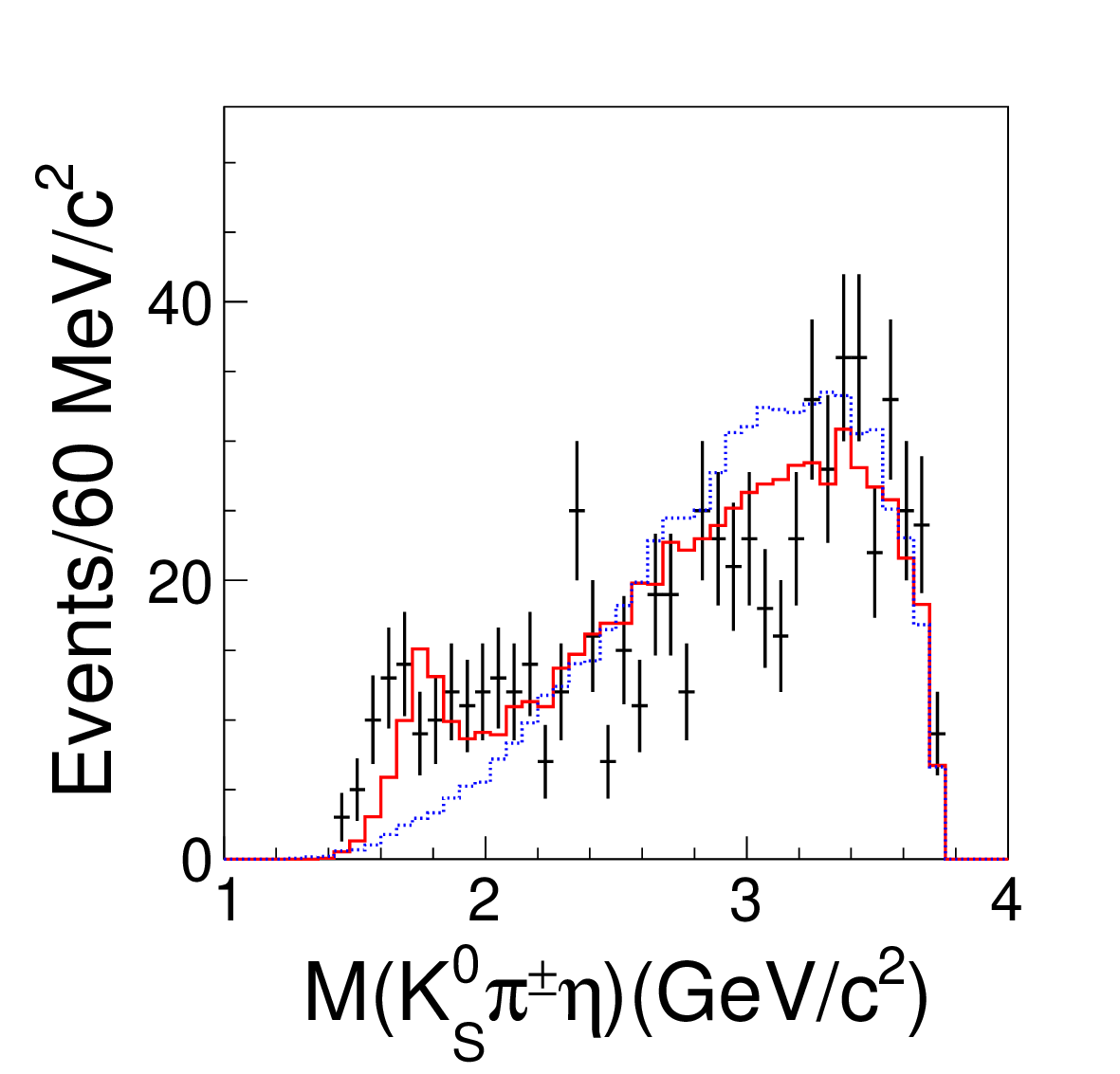}\put(-55,85){\bf(g) }
    }
\subfigure{
    \includegraphics[scale=0.2]{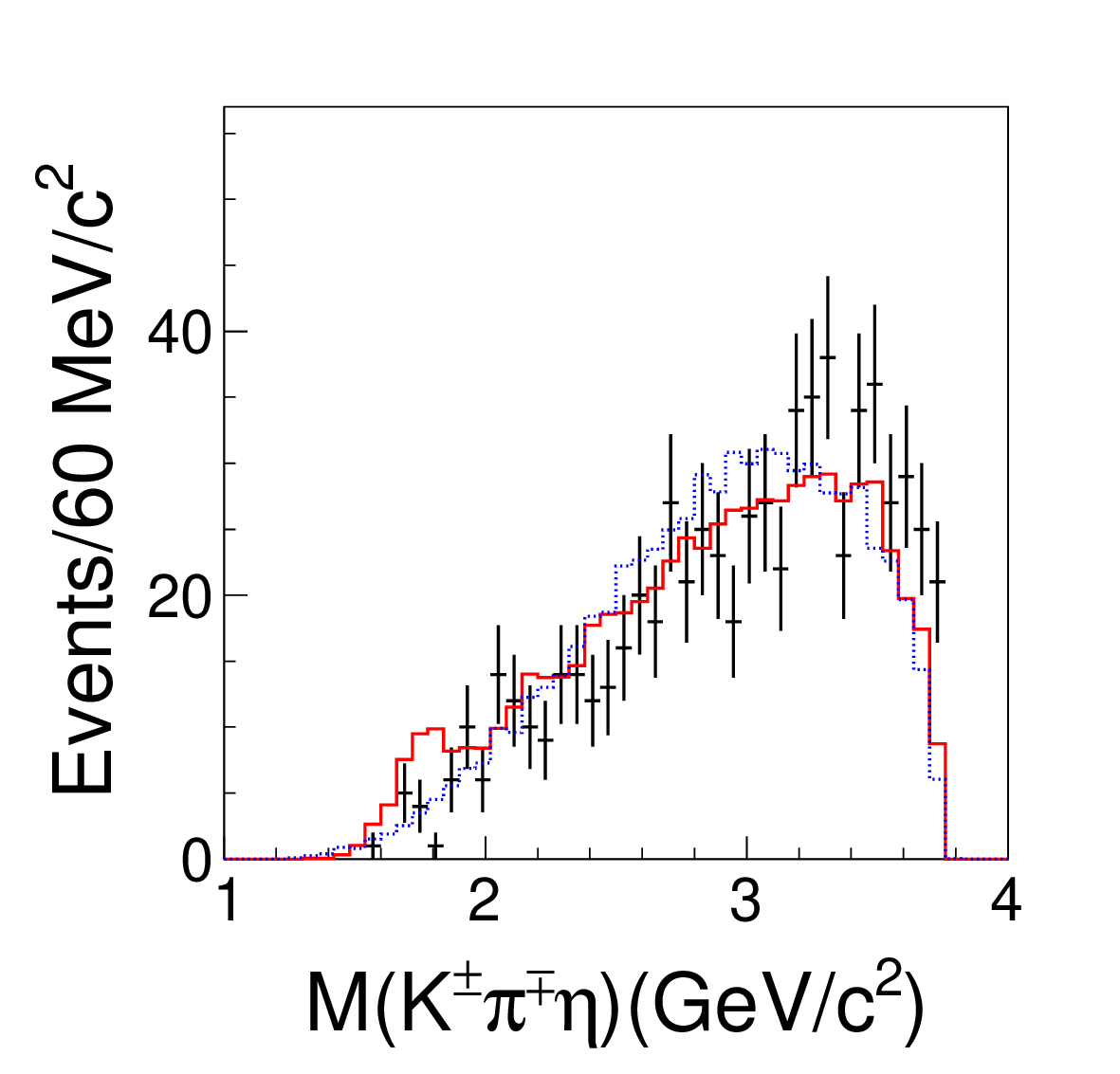}\put(-55,85){\bf(h) }
    }
\subfigure{
    \includegraphics[scale=0.2]{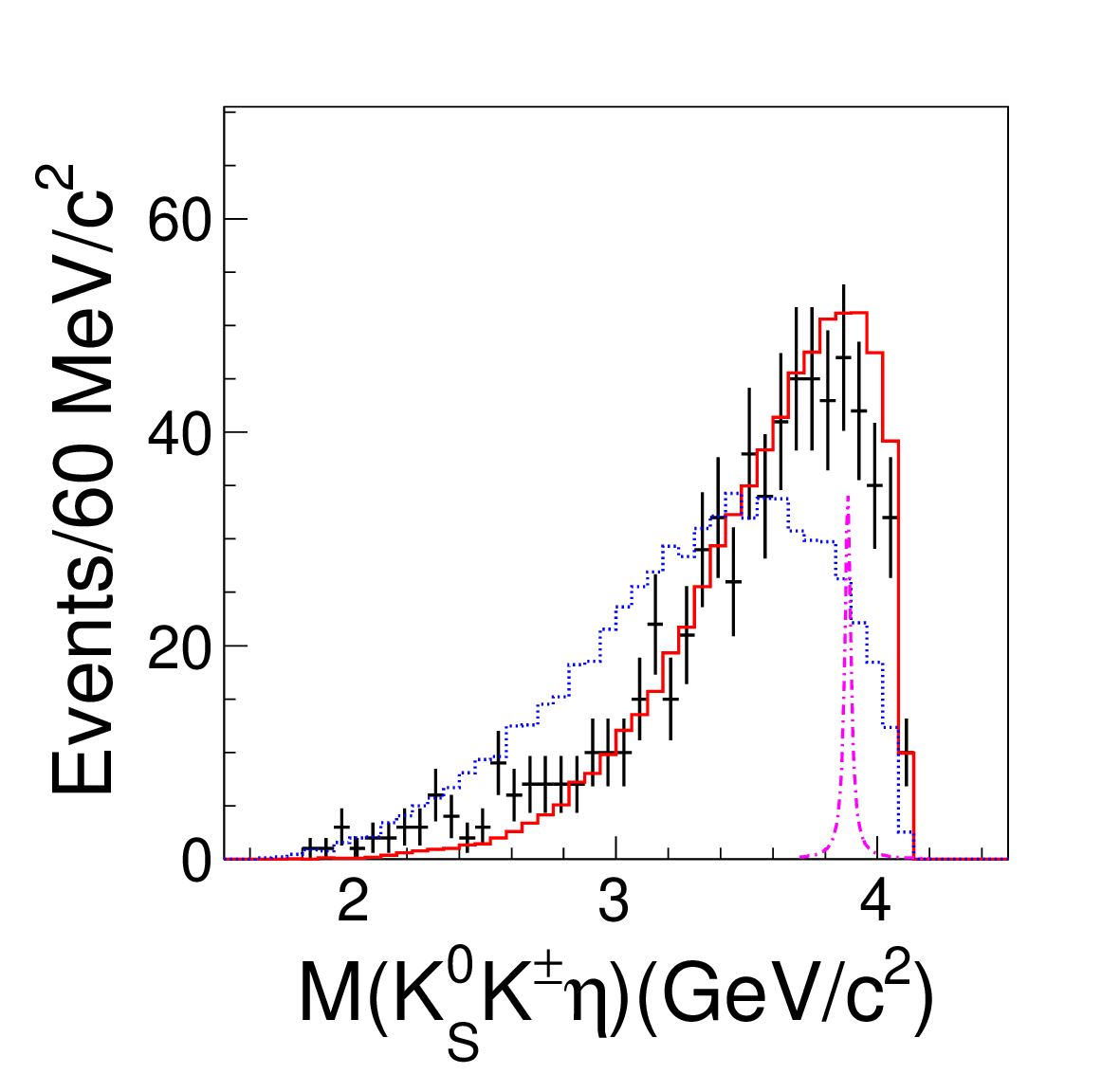}\put(-55,85){\bf(i) }
    }
\subfigure{
    \includegraphics[scale=0.2]{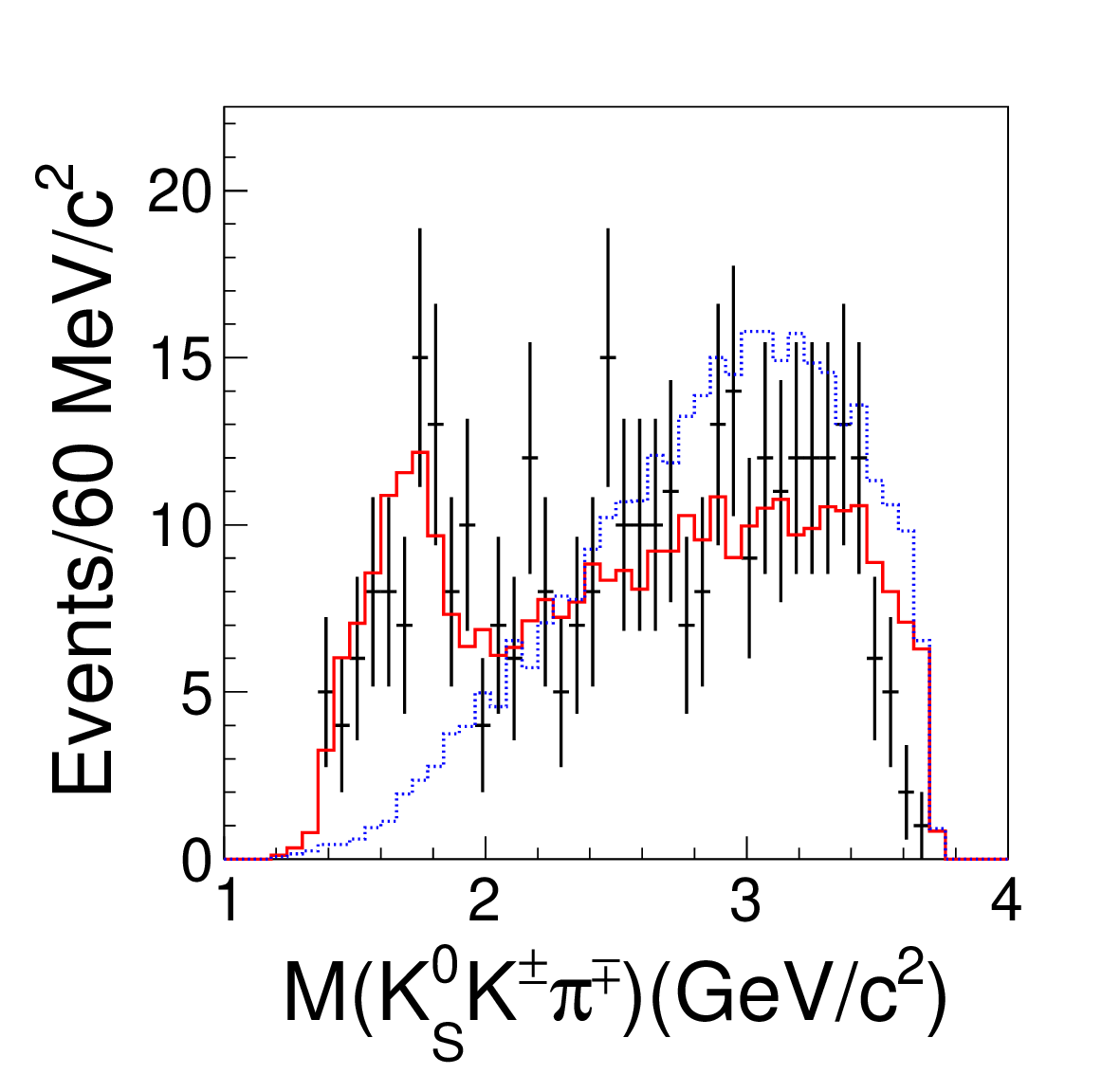}\put(-55,85){\bf(j) }
    }
\caption{(Color online) Invariant mass distributions of any two or three final state particles for the $\eta$ mode at 4.258 GeV. The black dots with error bars are the data. The red solid lines are the mixing MC sample. The blue dashed lines are the PHSP MC sample. The pink dash-dotted line in plot~(i) is the MC shape of the $Z_c(3900)$ with an arbitrary scale. \label{fig:resonances2}}
\end{figure}

The invariant mass distributions of any two or three final state particles at $\sqrt{s} = 4.258 ~\rm{GeV}$ are shown in
Figs.~\ref{fig:resonances1} and \ref{fig:resonances2} as examples. There are some intermediate states observed in this
four-body decay. To estimate the detection
efficiency, a data-driven method is implemented to produce an exclusive MC sample that more closely resembles the data.
This mixing MC sample includes intermediate resonances, such as $\rho(770)$ and $K^*(892)$, with couplings tuned
to approximately match those appear in the data sample and is weighted according to the momentum distributions
observed in the data sample. As illustrated in Figs.~\ref{fig:resonances1} and \ref{fig:resonances2}, the mixing MC sample
gives a much better description of the data than a phase space (PHSP) MC sample. The observed cross sections are presented
in Tables~\ref{tab:result_pi0} and \ref{tab:result_eta}, and illustrated in Fig.~\ref{lineshape}.

\begin{table*}[htbp]
\begin{center}
{\caption{Data sets and results of the Born cross section measurement for $e^+ e^- \to  K_S^0 K^\pm \pi^\mp \pi^0$. The table includes the integrated luminosity $\mathcal{L}$, the number of observed signals events $N_{\rm{sig}}$, the efficiency $\epsilon$, the ISR correction factor $(1+\delta^{\rm{ISR}})$, the vacuum polarization correction factor $\frac{1}{|1-\Pi(s)|^2}$, and the Born cross section $\sigma_{\rm{B}}$. The first errors are statistical and the second ones are systematic. The details of systematic uncertainties are described in Sec.~\ref{sec:syserr}.
} \label{tab:result_pi0}}
\begin{tabular}{ccrccccc}
\hline\hline
$\sqrt{s}~(\mathrm{GeV})$ &$\mathcal{L}~(\mathrm{pb^{-1}})$ & $N_{\rm{sig}}$~~ & $\epsilon~(\%)$& $(1+\delta^{\rm{ISR}})$&$\frac{1}{|1-\Pi(s)|^2}$&$\sigma_{\rm{B}}~(\mathrm{pb})$\\\hline
3.896&52.61&$469\pm22$ &16.76&1.03&1.05&74.41~$\pm$~3.47~$\pm$~3.35\\
4.008&481.96&$3335\pm58$ &16.41&1.05&1.04&58.02~$\pm$~1.01~$\pm$~2.61\\
4.086&52.63&$~307\pm18$ &16.70&1.06&1.05&47.52~$\pm$~2.73~$\pm$~2.14\\
4.189&43.09&$~240\pm16$ &16.31&1.08&1.06&45.38~$\pm$~2.94~$\pm$~2.04\\
4.208&54.55&$~269\pm17$ &15.49&1.11&1.06&40.95~$\pm$~2.50~$\pm$~1.84\\
4.217&54.13&$~257\pm16$ &16.03&1.11&1.06&38.28~$\pm$~2.40~$\pm$~1.72\\
4.226&1091.74&$5235\pm73$ &15.90&1.10&1.06&39.23~$\pm$~0.55~$\pm$~1.77\\
4.242&55.59&$~255\pm16$ &16.02&1.10&1.06&37.46~$\pm$~2.35~$\pm$~1.69\\
4.258&825.67&$3850\pm63$ &15.52&1.12&1.05&38.65~$\pm$~0.63~$\pm$~1.74\\
4.308&44.90&$~199\pm15$ &15.55&1.11&1.05&36.86~$\pm$~2.62~$\pm$~1.66\\
4.358&539.84&$~2167\pm47$ &15.38&1.12&1.05&33.53~$\pm$~0.72~$\pm$~1.51\\
4.387&55.18&$~237\pm16$ &16.00&1.15&1.05&33.68~$\pm$~2.20~$\pm$~1.52\\
4.416&1073.56&$3934\pm63$ &15.21&1.14&1.05&30.38~$\pm$~0.49~$\pm$~1.37\\
4.467&109.94&$~378\pm20$ &15.87&1.17&1.06&26.70~$\pm$~1.38~$\pm$~1.20\\
4.527&109.98&$~364\pm20$ &15.35&1.17&1.06&26.51~$\pm$~1.40~$\pm$~1.19\\
4.575&46.67&$~149\pm13$ &15.15&1.19&1.06&24.87~$\pm$~2.06~$\pm$~1.12\\
4.600&566.93&$1612\pm41$ &15.49&1.16&1.06&22.71~$\pm$~0.57~$\pm$~1.02\\
\hline\hline
\end{tabular}
\end{center}
\end{table*}

\begin{table*}[htbp]
\begin{center}
{\caption{Same as TABLE~\ref{tab:result_pi0} for $e^+ e^- \to  K_S^0 K^\pm \pi^\mp \eta$.
}\label{tab:result_eta}}
\begin{tabular}{cclcccc}
\hline\hline
$\sqrt{s}~(\mathrm{GeV})$ &$\mathcal{L}~(\mathrm{pb^{-1}})$ &$ ~~N_{\rm{sig}}$&$\epsilon~(\%)$&($1+\delta^{\rm{ISR}})$&$\frac{1}{|1-\Pi(s)|^2}$&$\sigma_{\rm{B}}~(\mathrm{pb})$\\\hline
3.896&52.61&~\,$76\pm9$ &18.22&1.02&1.05 &$27.23\pm3.22\pm1.26$\\
4.008&481.96&$516\pm24$&18.19&1.04&1.04  &$19.88\pm0.92\pm0.94$\\
4.085&52.63&$~\,42\pm7$ &18.07&1.05&1.05 &$14.71\pm2.45\pm0.70$\\
4.189&43.09&$~\,43\pm7$ &17.92&1.09&1.06 &$17.75\pm2.89\pm0.84$\\
4.208&54.55&$~\,43\pm7$ &17.76&1.08&1.06 &$14.20\pm2.31\pm0.61$\\
4.217&54.13&$~\,31\pm6$ &18.06&1.09&1.06 &$10.05\pm1.95\pm0.41$\\
4.226&1091.74&$942\pm31$&17.85&1.08&1.06 &$15.61\pm0.51\pm0.64$\\
4.242&55.59&$~\,45\pm7$ &17.86&1.08&1.06 &$14.63\pm2.28\pm0.60$\\
4.258&825.67&$655\pm26$&17.75&1.08&1.05  &$14.35\pm0.57\pm0.66$\\
4.308&44.90&$~\,32\pm6$ &17.59&1.12&1.05 &$12.67\pm2.38\pm0.55$\\
4.358&539.84&$349\pm19$&17.79&1.12&1.05  &$11.38\pm0.62\pm0.51$\\
4.387&55.18&$~\,40\pm6$ &17.44&1.11&1.05 &$13.05\pm1.96\pm0.62$\\
4.416&1073.56&$638\pm26$&17.56&1.11&1.05 &$10.62\pm0.43\pm0.49$\\
4.467&109.94&$~\,66\pm8$ &17.23&1.14&1.06&$10.62\pm1.29\pm0.52$\\
4.527&109.98&$~\,45\pm7$ &17.20&1.14&1.06&$~\,7.27\pm1.31\pm0.37$\\
4.575&47.67&$~\,27\pm5$ &17.29&1.15&1.06 &$~\,9.23\pm1.84\pm0.49$\\
4.600&566.93&$288\pm18$&17.20&1.18&1.06  &$~\,8.67\pm0.54\pm0.43$\\
\hline\hline
\end{tabular}
\end{center}
\end{table*}

\subsection{\boldmath Upper limits of $e^+e^-$ $\to$ $Y(4260)$ $\to$ $K_{S}^{0}$$K^{\pm}$$\pi^{\mp}$$\pi^0$ and $e^+e^-$ $\to$ $Y(4260)$ $\to$ $K_{S}^{0}$$K^{\pm}$$\pi^{\mp}$$\eta$}

\begin{figure}[htbp]
\subfigure{
    \includegraphics[width = 0.45\textwidth]{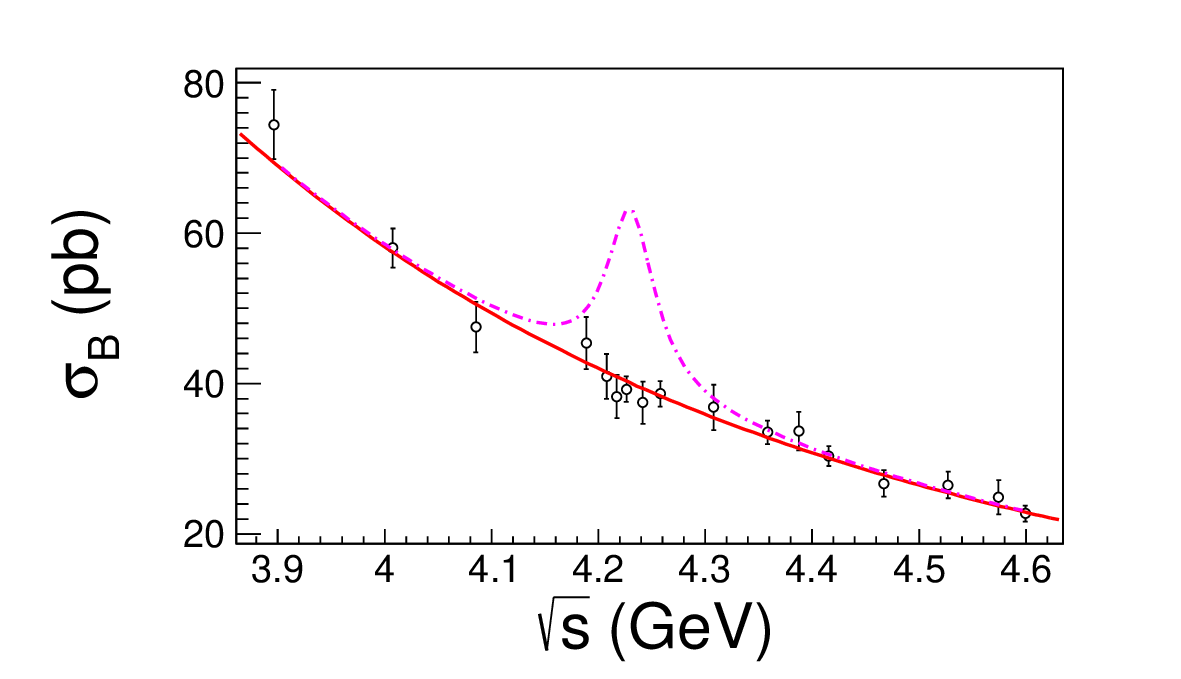}\put(-50,70){\bf(a)}
    }
\subfigure{
    \includegraphics[width = 0.45\textwidth]{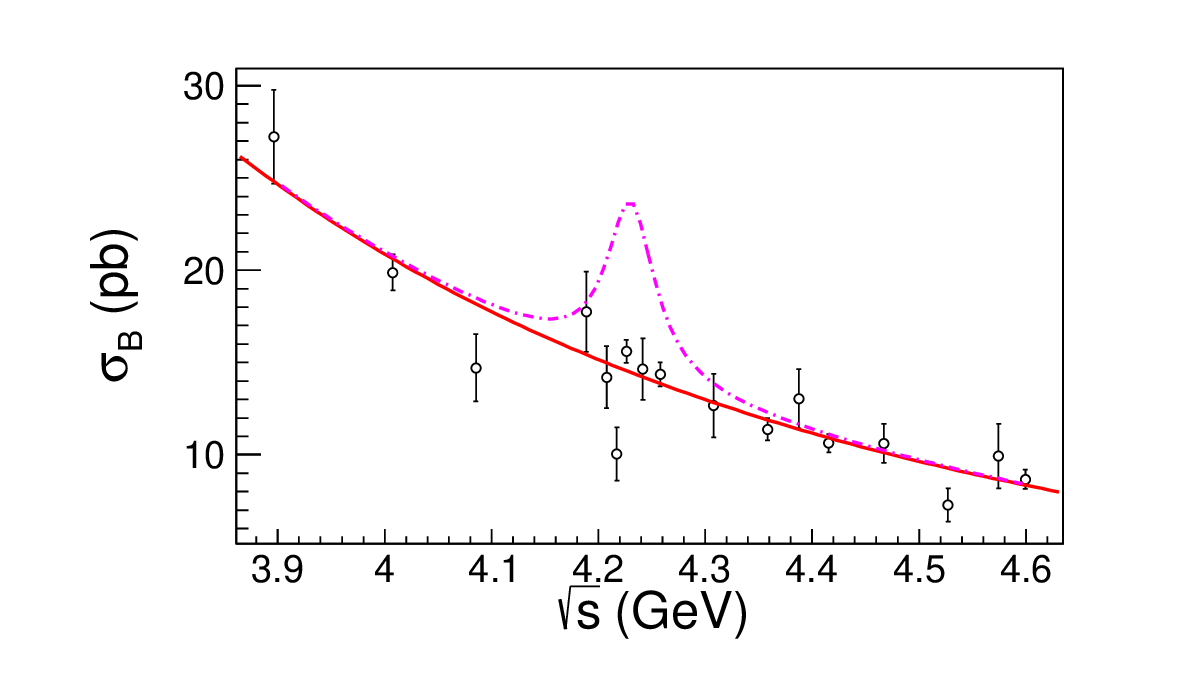}\put(-50,70){\bf(b)}
    }
\caption{(Color Online) Line shapes of Born cross sections for $e^+e^- \ra K_S^0K\pi\pi^0$ (a), and $e^+e^- \ra K_S^0K\pi\eta$ (b). The dots with error bars are the measured Born cross sections. The solid red lines are the fitted results with the function $f(\sqrt{s}) = \frac{p_0}{(\sqrt{s})^{p_1}}$ and parameters $p_0 = (6.14 \pm 1.54)\times10^6 ~\mathrm{(pb)}\cdot(\mathrm{GeV})^{p_1}$ and $p_1 = 6.68 \pm 0.17$ in the $\pi^0$ mode and $p_0 = (1.86 \pm 0.97)\times10^5~\mathrm(pb)\cdot\mathrm{(GeV)}^{p_1}$ and $p_1 = 6.56 \pm 0.36$ in the $\eta$ mode. The pink dash-dotted lines are the MC shape of the $Y(4260)$ with an arbitrary scale factor.}\label{lineshape}
\end{figure}

Since there is no obvious structure in the line shapes of the Born cross sections for $e^+ e^- \to
K_S^0 K^\pm \pi^\mp \pi^0$ and  $e^+ e^- \to K_S^0 K^\pm \pi^\mp \eta$, as shown in
Fig.~\ref{lineshape},
the upper limits of $Y(4260)$ $\to$ $K_S^0$ $K^\pm$ $\pi^\mp$ $\pi^0$ and $Y(4260)$ $\to$ $K_S^0$
$K^\pm$ $\pi^\mp$ $\eta$ are determined by fitting the line shapes with the function
$\sigma_{\rm{B}}(\sqrt{s}) =
c_0\cdot{f(\sqrt{s})} + \rm{BW(\sqrt{s})}$. Here $f(\sqrt{s})=\frac{p_0}{(\sqrt{s})^{p_1}}$ describes the continuum process $e^+ e^- \to K_S^0 K^\pm\pi^\mp\pi^0/\eta$, the parameters $p_0$ and $p_1$ are determined by fitting the line
shapes with only the continuum process. $\rm BW(\sqrt{s})$ given in Eq.~(\ref{bw})
\begin{equation}\label{bw}
\rm{BW(\sqrt{s})}=\frac{12\pi\Gamma_{\rm{e^+e^-}}\mathcal{B}\Gamma_{tot}}{(s-M^2)^2+M^2\Gamma_{tot}^2}
\end{equation}
is a relativistic Breit-Wigner
function describing the resonance $Y(4260)$, where $\rm M$, $\Gamma_{\rm{tot}}$, and $\Gamma_{\rm{e^+e^-}}$ are
the mass, full width, and electronic width of $Y(4260)$, respectively; $\mathcal{B}$ is the
branching fraction of the decay $Y(4260)$ $\to$ $K_S^0$ $K^\pm$ $\pi^\mp$ $\pi^0/\eta$.\\
\indent The mass and the full width of Y(4260) are set to the world average values $4230\pm8~\mathrm{MeV/c^2}$ and $55\pm 19~\mathrm{MeV/c^2}$~\cite{PDG}. The parameter $c_0$ is allowed to float during the fits, while the product
$\Gamma_{\rm{e^+e^-}}\mathcal{B}$ increases from $0$ to $0.5$
$\mathrm{eV}$ in step length of $0.001~\mathrm{eV}$. For each value of it, a fitting estimator
$Q^2$ defined  by Eq.~(\ref{q2})
 \begin{equation}\label{q2}
Q^{\rm{2}}=\sum\limits_{i}\frac{(\sigma_{\rm{B}_i}-h\cdot\sigma_{\rm{B}_i}^{\rm{fit}})^2}{\delta_{i}^{2}}+ \frac{(h-1)^{2}}{\delta_{c}^{2}}
\end{equation}
is obtained. Here
$\sigma_{\rm{B}}$ and $\sigma_{\rm{B}}^{\rm{fit}}$ are the measured and fitted Born cross sections, $\delta_i$ is the energy dependent part of the total uncertainty,
which includes
the statistical uncertainty and the energy dependent part of systematic uncertainty, the $\delta_c$ is the energy independent part of the systematic uncertainty (the systematic uncertainties are described in detail in Sec.~\ref{sec:syserr}),
$h$ is a free parameter introduced to take into account the correlation of different energy points, and the subscript $i$
indicates the index of each energy point~\cite{estimator} . The $Q^2$ is used to calculate the likelihood $L$ = $e^{-0.5Q^2}$, whose normalized distribution
is used to get the upper limits of $\Gamma_{\rm{e^+e^-}}\mathcal{B}$ at the $90\%$ confidence level (C.L.),
which is determined to be $0.050~\mathrm{eV}$ and $0.19~\mathrm{eV}$ for the $\pi^0$ mode
and the $\eta$ mode, respectively.

\subsection{\boldmath Upper limits on $\sigma_{\rm{B}}$$\big($$e^+e^-$ $\to$ $\pi^{0,\mp}$ $Z^{0,\pm}_c(3900)$, $Z^{0,\pm}_c(3900)$ $\to$ $K_{S}^{0}$$K^{\pm}$$\pi^{\mp,0}/\eta$$\big)$}

Since there is no obvious $Z_c(3900)$ signal in the invariant mass distributions of $K_S^0 K^{\pm} \pi^{\mp,0}$ ($\pi^0$ mode) and $K_S^0 K^{\pm} \eta$ ($\eta$
mode), as shown in Figs.~\ref{fig:resonances1} and \ref{fig:resonances2}, the upper limits at the $90\%$ C.L. for
the production cross section $\sigma_{\rm{B}}\big($$e^+ e^-$ $\to$ $\pi$ $Z_c(3900)\big)$, with
$Z_c(3900)$ $\to$ $K_S^0$ $K$ $\pi/\eta$ are determined with an unbinned maximum likelihood fit to the invariant mass of $K_S^0 K \pi/\eta$
in the range $(3.7,4.1)~\mathrm{GeV}/c^2$, at the five energy points $4.226$, $4.258$, $4.358$, $4.416$, and $4.600$ $\mathrm{GeV}$. The contribution of non-$K_S^0$ or non-$\pi^0/\eta$
backgrounds
is negligible.
In the fit, the $Z_c(3900)$ signal is described by
the MC simulated shape, and the mass and width of the $Z_c(3900)$ are set to theirs world average value
$3886.6\pm2.4$ $\mathrm{MeV}/c^2$ and $28.2\pm2.6$ $\mathrm{MeV}/c^2$~\cite{PDG}, respectively. The background is described by a second order polynomial function.
The normalized likelihood distribution of the Born cross section $L(\sigma_{\rm{B}})$ is determined by changing the number of signal events from
$0$ to $150$ with a step size of $1$. The upper limit (UL) at the $90\%$ C.L. is calculated by solving the equation
\begin{equation}\label{eq:UL}
    0.1 = \int_{\rm{UL}}^{\infty} L(\sigma_{\rm{B}})d\sigma_{\rm{B}}\;,
\end{equation}
The final upper limits are shown in Table~\ref{tab:UL_Zc3900}, where all of the systematic uncertainties have
been considered, the details of which are explained in Sec.~\ref{sec:syserr}.
The ratio
\begin{equation*}
R = \frac{\sigma_B(e^+ e^- \to \pi Z_c(3900)\to\pi K_S^0 K
\pi/\eta)}{\sigma_B(e^+ e^- \to \pi Z_c(3900)\to\pi \pi J/\psi)}
\end{equation*}
is also given in Table~\ref{tab:UL_Zc3900}, where the cross sections for $e^+ e^-$ $\to$ $\pi^{\mp }Z_c(3900)^{\pm}\to\pi^+\pi^- J/\psi$ and $e^+ e^-$ $\to$ $\pi^0 Z_c(3900)^0\to\pi^0 \pi^0 J/\psi$ are from Ref.~\cite{Collaboration:2017njt} and Ref.~\cite{PhysRevLett.115.112003}, respectively.

\begin{table}[htbp]
%\begin{small}
    {\caption {Upper limits on $\sigma_{\rm{B}}\big($$e^+e^-$ $\rightarrow$ $\pi$$Z_c(3900)$,\ $Z_c(3900)$$\to$$K_S^0$$K$$\pi/\eta\big)$, and its ratio~(R) to $\sigma_{\rm{B}}(e^+e^- \to \pi Z_c(3900),\ Z_c(3900) \to\pi J/\psi)$ at the 90\% C.L..}\label{tab:UL_Zc3900}}
    \begin{center}
    \begin{tabular}{cccc}\hline\hline
                                                              &$\sqrt{s}~(\rm{GeV})$& $ \sigma_{\rm{B}}~(\mathrm{pb}) $& R\\\hline
    \multirow{3}{*}{$e^+e^- \rightarrow \pi^0 Z_c(3900)^0,$}             &4.226&$<0.24$&$<2.5\times10^{-2}$\\
    \multirow{4}{*}{$Z_c(3900)^0 \to K_S^0 K^{\pm} \pi^\mp$}             &4.258&$<0.38$&$<1.2\times10^{-1}$\\
                                                                         &4.358&$<0.51$&$<2.6\times10^{-1}$\\
                                                                         &4.416&$<0.27$&-\\
                                                                         &4.600&$<0.33$&-\\\hline
    \multirow{3}{*}{$e^+e^- \rightarrow \pi^{\pm} Z_c(3900)^{\mp},$}     &4.226&$<0.17$&$<9.1\times10^{-3}$\\
    \multirow{4}{*}{$Z_c(3900)^{\mp} \to K_S^0 K^{\mp} \pi^0$}           &4.258&$<0.28$&$<5.6\times10^{-2}$\\
                                                                         &4.358&$<0.57$&-\\
                                                                         &4.416&$<0.34$&-\\
                                                                         &4.600&$<0.45$&-\\\hline
    \multirow{3}{*}{$e^+e^- \rightarrow \pi^{\pm} Z_c(3900)^{\mp},$}     &4.226&$<0.18$&$<1.0\times10^{-2}$\\
    \multirow{4}{*}{$Z_c(3900)^{\mp} \to K_S^0 K^{\mp} \eta$}            &4.258&$<0.56$&$<1.4\times10^{-1}$\\
                                                                         &4.358&$<0.53$&-\\
                                                                         &4.416&$<0.76$&-\\
                                                                         &4.600&$<0.58$&-\\\hline\hline
    \end{tabular}
    \end{center}
 %   \end{small}
\end{table}

\subsection{Systematic Uncertainties}\label{sec:syserr}
Various sources of systematic uncertainty are investigated in the $K_S^0 K^{\pm} \pi^{\mp} \pi^0$ and $K_S^0 K^{\pm} \pi^{\mp} \eta$
lineshape measurement.
We assume that the systematic uncertainties associated with
the physics model used in the MC simulation, the luminosity, tracking, PID, $\gamma$ reconstruction efficiency, $K^0_S$ reconstruction efficiency, ISR correction factor, vacuum polarization factor and quoted BFs are energy independent, while the other systematic effects are energy dependent.

For the $\pi^0$ mode, a data-driven MC method is developed to obtain the efficiency. To estimate the uncertainty of this
method, one thousand testing samples of $e^+ e^- \to K_S^0 K^{\pm} \pi^{\mp} \pi^0$ are generated with eighteen different
physics processes with random ratios, the ratio of each process is generated using uniform distribution between 0 to 1 and
then normalized by the summation of these eighteen ratios. The difference between the estimated and the real efficiencies
is fitted with a Gaussian function. The fit results give a mean of $0.4\%$ which is neglected, and a width of $0.9\%$ which is taken as the
systematic uncertainty from the data-driven MC method. For the $\eta$ mode, which has much lower statistics than the
$\pi^0$ mode, alternative mixing ratios are used to generate a new MC sample and the efficiency difference between the two MC samples is
adopted as the systematic uncertainty.

The uncertainty on the integrated luminosity is estimated to be $1.0\%$ using Bhabha events~\cite{Ablikim:2015nan}.

Both the uncertainties of tracking and PID for charged tracks originating at the interaction point are determined to be $1.0\%$ per track using $J/\psi \to K_S^0 K^\pm \pi^\mp$, $J/\psi \to p \bar{p} \pi^+ \pi^-$, and $J/\psi \to \pi^+ \pi^- \pi^0$~\cite{Ablikim:2011kv} as control samples.

The uncertainty due to photon reconstruction efficiency is $1.0\%$ per photon, which is derived from studies of $J/\psi \to \rho^0 \pi^0, \rho^0 \to \pi^+ \pi^-, \pi^0 \to \gamma \gamma$~\cite{pi0_rec_eff}.

The uncertainty associated with the $K_S^0$ reconstruction is studied using $J/\psi \rightarrow K^*(892)^\pm K^\mp, K^*(892)^\pm \to K_S^0 \pi^\pm$ and $J/\psi \rightarrow \phi K_S^0 K^\pm \pi^\mp$ control samples and is estimated to be 1.2\%~\cite{Ks_err}.

The ISR correction factor introduces a $1.0\%$ uncertainty since the termination condition of the recursion method used to get the correction factor is $1.0\%$ between the last two iterations.

The uncertainty due to the vacuum polarization factor is
%smaller than $0.1\%$ and ignored
found to be negligible~\cite{VP}. The uncertainties of the quoted BFs are also considered.

The energy dependent ones include the systematic uncertainties from the choosing about mass window and sideband regions of $K_S^0$, $\pi^0$,
and $\eta$ and the kinematic fit.
The uncertainties associated with the $K_S^0$, $\pi^0$, and $\eta$ invariant mass regions are determined by changing them from $(0.488, 0.508)$ to $(0.483, 0.513) ~\mathrm{GeV}/c^2$,
$(0.12, 0.15)$ to $(0.115, 0.155) ~\mathrm{GeV}/c^2$ and $(0.52,0.58)$ to $(0.51,0.59) ~\mathrm{GeV}/c^2$ for the $K_S^0$, $\pi^0$ and $\eta$, respectively. The
differences in the efficiencies are taken as the corresponding systematic uncertainties.

The uncertainties due to the side-band regions are determined by changing the side-band region to $M_{\eta} \in (0.44,0.47)\cup(0.63,0.66)~\mathrm{GeV}/c^2$, $M_{\pi^0} \in
(0.08,0.095)\cup(0.175,0.19)~\mathrm{GeV}/c^2$ and $M_{K_S^0} \in (0.463,0.473)\cup(0.523,0.533)~\mathrm{GeV}/c^2$. The differences are taken as the associated systematic uncertainties.

The uncertainty associated with the kinematic fit is determined by comparing the efficiencies with and without corrections to the track helix parameters~\cite{Ablikim:2012pg}.

Assuming all sources of systematic uncertainties are independent, the total uncertainties are the sums of the individual values in quadrature
(Table~\ref{syserr}).

\begin{table*}[htbp]{\caption{Summary of systematic uncertainties (in \%).}\label{syserr}}
\begin{center}
%\hskip -0.5cm

\begin{tabular}{l|clllllllllllllllll}\hline\hline
&$\sqrt{s}~(\rm{GeV})$ & $3.896$&$4.008$ &$4.085$ &$4.189$ &$4.208$ &$4.217$ &$4.226$ &$4.242$ &$4.258$ &$4.308$ &$4.358$ &$4.387$ &$4.416$ &$4.467$ &$4.527$ &$4.575$ &$4.600$\\\hline
\multirow{7}{*}{\begin{sideways}both mode\end{sideways}}
&$\mathcal{L}$& 1.0      &1.0       &1.0      &1.0      &1.0      &1.0      &1.0      &1.0      &1.0      &1.0  &1.0      &1.0      &1.0      &1.0      &1.0      &1.0      &1.0    \\
&$K_S^0$ reconstruction           & 1.2    &1.2     &1.2    &1.2    &1.2    &1.2    &1.2    &1.2    &1.2    &1.2    &1.2    &1.2    &1.2    &1.2    &1.2  &1.2    &1.2  \\
&Tracking                  & 2.0      &2.0       &2.0      &2.0      &2.0      &2.0      &2.0      &2.0      &2.0      &2.0      &2.0      &2.0      &2.0      &2.0      &2.0      &2.0      &2.0    \\
&PID                    & 2.0      &2.0       &2.0      &2.0      &2.0      &2.0      &2.0      &2.0      &2.0      &2.0      &2.0      &2.0      &2.0      &2.0      &2.0      &2.0      &2.0    \\
&$\gamma$ reconstruction  & 2.0      &2.0       &2.0      &2.0      &2.0      &2.0      &2.0      &2.0      &2.0      &2.0      &2.0      &2.0      &2.0      &2.0      &2.0      &2.0      &2.0    \\
&$\rm{BF}_{K_S^0 \rightarrow \pi^+ \pi^-}$    & 0.1   &0.1    &0.1   &0.1   &0.1   &0.1   &0.1   &0.1   &0.1   &0.1   &0.1   &0.1   &0.1   &0.1   &0.1   &0.1   &0.1 \\
&$(1+\delta^{ISR})$     & 1.0   &1.0    &1.0   &1.0   &1.0   &1.0   &1.0   &1.0   &1.0   &1.0   &1.0   &1.0   &1.0   &1.0   &1.0 &1.0   &1.0 \\
\hline\multirow{7}{*}{\begin{sideways}$\pi^0$ mode\end{sideways}}
&Mixing MC            & 0.9      &0.9       &0.9      &0.9      &0.9      &0.9      &0.9      &0.9      &0.9      &0.9      &0.9      &0.9      &0.9      &0.9      &0.9      &0.9      &0.9    \\
&Kinematic fit                                     & 0.3   &0.3    &0.3   &0.3   &0.3   &0.4   &0.3   &0.3   &0.4   &0.1   &0.2   &0.2   &0.2   &0.4   &0.4   &0.3   &0.4 \\
&$\pi^0$ mass interval   & 0.6   &0.6    &0.6   &0.6   &0.6   &0.6   &0.6   &0.6   &0.5   &0.5   &0.6   &0.6   &0.6   &0.6   &0.8   &0.8   &0.8 \\

&$K_S^0$ mass interval  &0.1&	0.1& 0.1& 0.1& 0.1& 0.1& 0.1& 0.1& 0.3& 0.3& 0.2& 0.3& 0.3& 0.3& 0.2& 0.2& 0.2\\

&side-band           & 0.2   &0.2    &0.2   &0.2   &0.1   &0.1   &0.1   &0.1   &0.1   &0.1   &0.1   &0.1   &0.1   &0.1   &0.1   &0.1   &0.1 \\

&$\rm{BF}_{\pi^0 \rightarrow \gamma \gamma}$  & 0.1  &0.1   &0.1  &0.1  &0.1  &0.1  &0.1  &0.1  &0.1  &0.1 &0.1   &0.1  &0.1  &0.1  &0.1  &0.1  &0.1\\

%&Total                    & 4.1  &4.1    &4.1   &4.1   &4.1   &4.1   &4.1   &4.1   &4.1   &4.1   &4.1   &4.1   &4.1   &4.1   &4.1   &4.2   &4.2 \\
&Total                    & 4.1  &4.1    &4.1   &4.1   &4.1   &4.1   &4.1   &4.1   &4.1   &4.1   &4.1   &4.1   &4.1   &4.1   &4.1   &4.1   &4.1 \\
\hline\multirow{7}{*}{\begin{sideways}$\eta$ mode\end{sideways}}
&Mixing MC &0.2 & 1.4 & 1.1 & 1.0 & 1.2 & 0.3 & 0.1 & 0.4 & 1.6 & 0.5 & 1.4 & 1.2 & 0.3 & 1.6 & 1.5 & 0.6 & 0.9\\
&Kinematic fit    & 0.3   &0.2    &0.2   &0.2   &0.2   &0.2   &0.2   &0.3   &0.3   &0.2   &0.1   &0.3   &0.4   &0.1   &0.2   &0.2   &0.2 \\
&$\eta$ mass interval    &1.6&	1.6&	1.6&	1.6&	0.6&	0.6&	0.6&	0.6&	0.9&	0.9&	1.0&	0.4&	0.4&	0.4&	1.7&	1.7&	1.7\\
&$K_S^0$ mass interval  &1.7&	1.7& 1.7&	1.7&	0.7&	0.7& 0.7& 0.7& 1.4& 1.4& 1.1& 2.1& 2.1& 2.1& 2.3& 2.3& 2.3\\
&side-band &  0.1&	0.1&	0.1&	0.1&	0.6&	0.6&	0.6&	0.6&	0.5&	0.5&	0.4&	1.1&	1.1&	1.1&	0.4&	0.4& 0.4\\
&$\rm{BF}_{\eta \rightarrow \gamma \gamma}$  & 0.5  &0.5   &0.5  &0.5  &0.5  &0.5  &0.5  &0.5  &0.5  &0.5 &0.5   &0.5  &0.5  &0.5  &0.5  &0.5  &0.5\\
&Total &              4.6& 4.8& 4.7& 4.7& 4.3& 4.1& 4.1& 4.1& 4.6& 4.4& 4.5& 4.8& 4.7& 4.9& 5.1& 4.9& 5.0\\
\hline\hline
\end{tabular}
\end{center}
\end{table*}

The systematic uncertainties that affect the upper limits on $\sigma_{\rm{B}}\big(e^+ e^- \to \pi Z_c(3900), Z_c(3900) \to K_S^0 K \pi/\eta\big)$ are considered in two categories: multiplicative and non-multiplicative. The non-multiplicative systematic uncertainties on the signal shape and the background shape are considered by changing
the signal shape to a Breit-Wigner function and varying the fit range, the parameters of the $Z_c(3900)$, and the order of the polynomial functions in the
fit. The maximum upper limits are
adopted for all combinations of these variations. The intermediate states in the $Z_c(3900)$ decay are considered by generating signal MC samples with alternative
processes $Z_c(3900) \to K^*(892) K$, $K^*(892) \to K (K_S^0) \pi$ ($\pi^0$ mode), and  $Z_c(3900) \to a_0(980) \eta$,  $a_0(980) \to  K_S^0 K$ ($\eta$ mode). The
efficiency difference is considered as a multiplicative systematic
uncertainty. All of the systematic uncertainties, which are listed in Table~\ref{syserr}, excluding the side-band item and mixing MC item,  are considered as the multiplicative
systematic uncertainties. The effects of multiplicative systematic uncertainties are taken into account by convolving the distribution of $L(\sigma_{\rm{B}})$ with
a probability distribution function of sensitivity~$(S)$, which is assumed to be a Gaussian function with central
value $\hat{S}$ and standard deviation $\delta_S$~\cite{smear}:
\begin{equation}
    L^\prime(\sigma_{\rm{B}})=\int_{0}^{1}L(\frac{S}{\hat{S}}\sigma_{\rm{B}})\cdot e^{-\frac{(S-\hat{S})^2}{2\delta_{\sc{s}}^2}}dS \;.
    \end{equation}
Here $S$ is the sensitivity that refers to the denominator of Eq.~(\ref{eq:sigma}) and $\delta_{\sc{ s}}$ is the total multiplicative systematic uncertainty. $L^\prime(\sigma_{\rm{B}})$ is the likelihood distribution of the Born cross section after the multiplicative systematic uncertainties are incorporated.

\section{Summary}
The Born cross sections for $e^+e^- \to K_S^0 K^\pm \pi^\mp \pi^0$ and $K_S^0 K^\pm \pi^\mp \eta$ are
measured with data samples collected at center-of-mass energies from $3.90$ to
$4.60~\mathrm{GeV}$. Since no clear structure is observed, the upper limits of the product
$\Gamma_{\rm{e^+e^-}}\mathcal{B}$ $\big($ $Y(4260)$
$\to$$K_S^0$$K^\pm$$\pi^\mp$$\pi^0$$\big)$ at $90\%$ C.L.
is estimated to be less than $0.05~\mathrm{eV} $ and that of $\Gamma_{\rm {e^+e^-}}
{\mathcal{B}}\big(Y(4260)\to$$K_S^0$$K^\pm$$\pi^\mp$$\eta$$\big)$ is
estimated to be smaller than $0.19~\mathrm{eV}$. Ref.~\cite{Ablikim:2016qzw} reported four
solutions of the product $\Gamma_{\rm{e^+e^-}}\mathcal{B}\big{(}Y(4260)\to$
$\pi^+\pi^-J/\psi\big{)}$, in which the maximum is $13.3\pm1.4~\mathrm{eV}$ and the minimum is
$1.5\pm0.3~\mathrm{eV}$. Comparing them with our results, the branching fraction of the $Y(4260)$ decaying into $K_S^0 K^{\pm}
\pi^{\mp} \pi^0$ and $K_S^0 K^{\pm}
\pi^{\mp} \eta$ is much smaller, which indicates a
much smaller coupling of the $Y(4260)$ to the light hadrons $K_S^0$$K^{\pm}$$\pi^{\mp}$$\pi^0$
and $K_S^0$$K^{\pm}$$\pi^{\mp}$$\eta$. We
also search for $e^+e^- \to \pi Z_c(3900),\ Z_c(3900) \to K_S^0 K
\pi/\eta$ and no obvious $Z_c(3900)$ signal is observed in the charged nor neutral mode. The 90\%
C.L. upper limits on the cross sections are given at $\sqrt{s}=$ $4.226$, $4.258$, $4.358$, $4.416$, and $4.600~\mathrm{GeV}$.
The absence of a signal suggests that the cross sections for light hadron decay modes are small and
that the annihilation of $c\bar{c}$ in the $Y(4260)$ and $Z_c(3900)$ is suppressed. Additional
exploration of light hadron decay modes is needed to confirm the hypotheses.
\acknowledgements
The BESIII collaboration thanks the staff of BEPCII and the IHEP computing center for their strong support. This work is supported in part by National Key Basic Research Program of China under Contract No. 2015CB856700; National Natural Science Foundation of China (NSFC) under Contracts Nos. 11335008, 11425524, 11625523, 11635010, 11735014; the Chinese Academy of Sciences (CAS) Large-Scale Scientific Facility Program; the CAS Center for Excellence in Particle Physics (CCEPP); Joint Large-Scale Scientific Facility Funds of the NSFC and CAS under Contracts Nos. U1532257, U1532258, U1732263; CAS Key Research Program of Frontier Sciences under Contracts Nos. QYZDJ-SSW-SLH003, QYZDJ-SSW-SLH040; 100 Talents Program of CAS; INPAC and Shanghai Key Laboratory for Particle Physics and Cosmology; German Research Foundation DFG under Contracts Nos. Collaborative Research Center CRC 1044, FOR 2359; Istituto Nazionale di Fisica Nucleare, Italy; Koninklijke Nederlandse Akademie van Wetenschappen (KNAW) under Contract No. 530-4CDP03; Ministry of Development of Turkey under Contract No. DPT2006K-120470; National Science and Technology fund; The Swedish Research Council; U. S. Department of Energy under Contracts Nos. DE-FG02-05ER41374, DE-SC-0010118, DE-SC-0010504, DE-SC-0012069; University of Groningen (RuG) and the Helmholtzzentrum fuer Schwerionenforschung GmbH (GSI), Darmstadt


\begin{thebibliography}{9}
\bibitem{XYZ_review1}
H. X. Chen, W. Chen, X. Liu, and S. L. Zhu,
\href{https://doi.org/10.1016/j.physrep.2016.05.004}{Phys.\ Rep.\ {\bf 639}, 1 (2016)}.
\bibitem{XYZ_review2}
A. Ali, J. S. Lange, and S. Stone,
\href{https://doi.org/10.1016/j.ppnp.2017.08.003}
{Prog. Part. Nucl. Phys. {\bf 97}, 123 (2017).}
\bibitem{PhysRevLett.95.142001}
  B.~Aubert {\it et al.} (BABAR Collaboration),
  %``Observation of a broad structure in the $\pi^+ \pi^- J/\psi$ mass spectrum around 4.26-GeV/c$^2$,''
  \href{https://doi.org/10.1103/PhysRevLett.95.142001}
  {Phys.\ Rev.\ Lett.\  {\bf 95}, 142001 (2005).}
  %doi:10.1103/PhysRevLett.95.142001
  %[hep-ex/0506081].
  %%CITATION = doi:10.1103/PhysRevLett.95.142001;%%
  %633 citations counted in INSPIRE as of 31 Aug 2016
\bibitem{Y4260_pi0jpsi}
T. E. Coan {\it et al.} (CLEO Collaboration),
\href{https://doi.org/10.1103/PhysRevLett.96.162003}
{Phys.\ Rev.\ Lett.\ {\bf 96}, 162003 (2006).}
% arXiv:hep-ex/0602034.


%\cite{Li:2018fdr}
%\bibitem{Li:2018fdr}
\bibitem{Y4260_piddstar}
  K.~Li. (BESIII Collaboration),
  %``New results from $Y(4260)$ decays at BESIII,''
  \href{https://doi.org/10.22323/1.310.0108}
  {PoS Hadron 2017 {\bf 108}, (2018).}
  %doi:10.22323/1.310.0108
  %%CITATION = doi:10.22323/1.310.0108;%%

\bibitem{Y4260notseen1} B. Aubert, {\it et al.} (BABAR Collaboration),
%Study of the Exclusive Initial-State Radiation Production of the D¡¥D System,
\href{https://doi.org/10.1103/PhysRevD.76.111105}{
Phys.\ Rev.\ D {\bf 76}, 111105 (2007).}
%arXiv:hep-ex/0607083, doi:10.1103/PhysRevD.76.111105.
%2

\bibitem{Y4260notseen2}
 G. Pakhlova {\it et al.} (Belle Collaboration),
%Measurement of the near-threshold e+e?? ! D()D() cross section using initial-state radiation,
\href{https://doi.org/10.1103/PhysRevLett.98.092001}
{Phys.\ Rev.\ Lett.\ {\bf 98}, 092001 (2007).}
%arXiv:hep-ex/0608018, doi:10.1103/PhysRevLett.98.092001.

\bibitem{Y4260notseen3}
G. Pakhlova {\it et al.} (Belle Collaboration),
%Observation of (4415) ! D¡¥D 2(2460) decay using initial-state radiation,
 \href{https://doi.org/10.1103/PhysRevLett.100.062001}
 {Phys.\ Rev.\ Lett.\ {\bf 100}, 062001 (2008).}
 % arXiv:0708.3313, doi:10.1103/PhysRevLett.100.062001.
\bibitem{Y4260notseen4}
G. Pakhlova {\it et al.} (Belle Collaboration),
% Measurement of the e+e?? ! D0D??+ cross section using initial-state radiation,
\href{https://doi.org/10.1103/PhysRevD.80.091101}
{Phys.\ Rev.\ D \ {\bf 80}, 091101 (2009).}
% arXiv:0908.0231, doi:10.1103/PhysRevD.80.091101.
\bibitem{Y4260notseen5}
D. Cronin-Hennessy {\it et al.} (CLEO Collaboration),
%Measurement of Charm Production Cross Sections in e+e?? Annihilation at Energies between 3.97 and 4.26 GeV,
\href{https://doi.org/10.1103/PhysRevD.80.072001}
{Phys.\ Rev.\ D\ {\bf 80}, 072001 (2009).}
%arXiv:0801.3418, doi:10.1103/PhysRevD.80.072001.
\bibitem{Y4260notseen6}
B. Aubert {\it et al.} (BABAR Collaboration),
% Exclusive Initial-State-Radiation Production of the D¡¥D, D ¡¥D, and D ¡¥D Systems,
\href{https://doi.org/10.1103/PhysRevD.79.092001}
{Phys.\ Rev.\ D\ {\bf 79}, 092001 (2009).}
%arXiv:0903.1597, doi:10.1103/PhysRevD.79.092001.
\bibitem{Y4260notseen7}
P. del Amo Sanchez {\it et al.} (BABAR Collaboration),
 %Exclusive Production of D+s D??s , D+s D??s , and D+s D??s via e+e?? Annihilation with Initial-State-Radiation,
\href{https://doi.org/10.1103/PhysRevD.82.052004}
{Phys.\ Rev.\ D\ {\bf 82}, 052004 (2010).}
 %arXiv:1008.0338, doi:10.1103/PhysRevD.82.052004.

\bibitem{Y4260notseen8}
Z. Q. Liu, X. S. Qin, and C. Z. Yuan,
%Combined fit to BaBar and Belle data on e+e?? ! +?? (2S ),
\href{https://doi.org/10.1103/PhysRevD.78.014032}
{Phys.\ Rev.\ D\ {\bf 78}, 014032 (2008).}
%arXiv:0805.3560, doi:10.1103/PhysRevD.78.014032.
\bibitem{Y4260notseen9}
X. L. Wang {\it et al.} (Belle Collaboration),
%Observation of (4040) and (4160) decay into J= ,
\href{https://doi.org/10.1103/PhysRevD.87.051101}
{Phys.\ Rev.\ D\ {\bf 87}, 051101 (2013).}
% arXiv:1210.7550, doi:10.1103/PhysRevD.87.051101.
\bibitem{Y4260notseen10}
B. Aubert {\it et al.} (BABAR Collaboration),
%A Structure at 2175-MeV in e+e?? !  f0(980) Observed via Initial-State Radiation,
\href{https://doi.org/10.1103/PhysRevD.74.091103}
{Phys.\ Rev.\ D\ {\bf 74}, 091103 (2006).}
%arXiv:hep-ex/0610018, doi:10.1103/PhysRevD.74.091103.
\bibitem{Y4260notseen11}
B. Aubert {\it et al.} (BABAR Collaboration),
%Measurements of e+e?? ! K+K??, K+K??0 and K0s K cross- sections using initial state radiation events,
\href{https://doi.org/10.1103/PhysRevD.77.092002}
{Phys.\ Rev.\ D\ {\bf 77}, 092002 (2008).}
% arXiv:0710.4451, doi:10.1103/PhysRevD.77.092002.
\bibitem{Y4260notseen12}
B. Aubert {\it et al.} (BABAR Collaboration),
%A Study of e+e?? ! p ¡¥p using initial state radiation with BABAR,
\href{https://doi.org/10.1103/PhysRevD.73.012005}
{Phys.\ Rev.\ D\ {\bf 73}, 012005 (2006).}
%arXiv:hep-ex/0512023, doi:10.1103/PhysRevD.73.012005.

\bibitem{Y4260_4S_1}
F. J. Llanes-Estrada,
% Y(4260) and possible charmonium assignment,
\href{https://doi.org/10.1103/PhysRevD.72.031503}
{Phys.\ Rev.\ D\ {\bf 72}, 031503 (2005).}
%arXiv:hep-ph/0507035, doi:
%10.1103/PhysRevD.72.031503.
\bibitem{Y4260_4S_2}
M. Shah, A. Parmar, and P. C. Vinodkumar,
\href{https://doi.org/10.1103/PhysRevD.86.034015}
%Leptonic and Digamma decay Properties of S-wave quarkonia states,
{Phys.\ Rev.\ D\ {\bf 86},  034015 (2012).}
%arXiv:1203.6184, doi:10.1103/PhysRevD.86.034015.
\bibitem{Y4260_4S_3}
B. Q. Li and K. T. Chao,
%Higher Charmonia and X,Y,Z states with Screened Potential,
\href{https://doi.org/10.1103/PhysRevD.79.094004}
{Phys.\ Rev.\ D\ {\bf 79}, 094004 (2009).}
% arXiv:0903.5506,
%doi:10.1103/PhysRevD.79.094004.
\bibitem{Y4260_3D}
A. Zhang,
%Charmonium spectrum and new observed states,
\href{https://doi.org/10.1016/j.physletb.2007.01.062}
{Phys.\ Lett.\ B\ {\bf 647}, 140 (2007).}
%arXiv:hep-ph/0603093, doi:10.1016/j.physletb.2007.01.062.

\bibitem{Y4260_hybrid_1}
S. L. Zhu,
%The Possible interpretations of Y(4260),
\href{https://doi.org/10.1016/j.physletb.2005.08.068}
{Phys.\ Lett.\ B\ {\bf 625}, 212 (2005).}
%arXiv:hep-ph/0507025, doi:10.1016/j.physletb.2005.08.068.
\bibitem{Y4260_hybrid_2}
E. Kou and O. Pene,
%Suppressed decay into open charm for the Y(4260) being an hybrid,
\href{https://doi.org/10.1016/j.physletb.2005.09.013}
{Phys.\ Lett.\ B\ {\bf 631}, 164 (2005).}
%arXiv:hep-ph/0507119, doi:10.1016/j.physletb.2005.09.013.
\bibitem{Y4260_4q_1}
L. Maiani, F. Piccinini, A. D. Polosa, and V. Riquer,
%Four quark interpretation of Y(4260),
\href{https://doi.org/10.1103/PhysRevD.72.031502}
{Phys.\ Rev.\ D\ {\bf 72}, 031502 (2005).}
% arXiv:hep-ph/
%0507062, doi:10.1103/PhysRevD.72.031502.
\bibitem{Y4260_4q_2}
N. V. Drenska, R. Faccini, and A. D. Polosa,
%Exotic Hadrons with Hidden Charm and Strangeness,
\href{https://doi.org/10.1103/PhysRevD.79.077502}
{Phys.\ Rev.\ D\ {\bf 79}, 077502 (2009).}
%arXiv:0902.2803, doi:10.1103/PhysRevD.79.077502
%\cite{PhysRevLett.99.182004}

\bibitem{Y4260_4q_3}
D. Ebert, R. N. Faustov, and V. O. Galkin,
%Masses of heavy tetraquarks in the relativistic quark model,
\href{https://doi.org/10.1016/j.physletb.2006.01.026}
{Phys.\ Lett.\ B\ {\bf 634}, 214 (2006).}
%arXiv:hep-ph/0512230, doi:10.1016/j.physletb.2006.01.026.
\bibitem{Y4260_4q_4}
D. Ebert, R. N. Faustov, and V. O. Galkin,
%Excited heavy tetraquarks with hidden charm,
\href{https://doi.org/10.1140/epjc/s10052-008-0754-8}
{Eur.\ Phys.\ J.\ C\ {\bf 58}, 399 (2008).}
% arXiv:0808.3912,doi:10.1140/epjc/s10052-008-0754-8.
\bibitem{Y4260_m_1}
C. Z. Yuan, P. Wang, and X. H. Mo,
%The Y(4260) as an !c1 molecular state,
\href{https://doi.org/10.1016/j.physletb.2006.01.031}
{Phys.\ Lett.\ B\ {\bf 634}, 399 (2006).}
%arXiv:hep-ph/0511107,doi:10.1016/j.physletb.2006.01.031.
\bibitem{Y4260_m_2}
G. J. Ding,
%Are Y(4260) and Z+(4248) are D1 ¡¥D or D0 ¡¥D Hadronic Molecules?,
\href{https://doi.org/10.1103/PhysRevD.79.014001}
{Phys.\ Rev.\ D\ {\bf 79}, 014001 (2009).}
%arXiv:0809.4818,doi:10.1103/PhysRevD.79.014001.
\bibitem{Y4260_m_3}
F. Close and C. Downum,
%On the possibility of Deeply Bound Hadronic Molecules from single Pion Exchange,
\href{https://doi.org/10.1103/PhysRevLett.102.242003}
{Phys.\ Rev.\ Lett.\ {\bf 102}, 242003 (2009).}
%arXiv:0905.2687, doi:10.1103/PhysRevLett.102.242003
\bibitem{Y4260_m_4}
F. Close, C. Downum, and C. E. Thomas,
%Novel Charmonium and Bottomonium Spectroscopies due to Deeply Bound Hadronic Molecules from Single Pion Exchange,
\href{https://doi.org/10.1103/PhysRevD.81.074033}
{Phys.\ Rev.\ D\ {\bf 81}, 074033 (2010).}
%arXiv:1001.2553, doi:10.1103/PhysRevD.81.074033.


\bibitem{Y4260_m_5}
M. Cleven, Q. Wang, F. K. Guo, C. Hanhart, U. G. Meissner, and Q. Zhao,
%Y(4260) as the first S -wave open charm vector molecular state?,
\href{https://doi.org/10.1103/PhysRevD.90.074039}
{Phys.\ Rev.\ D\ {\bf 90}, 074039 (2014).}
%arXiv:1310.2190, doi:10.1103/PhysRevD.90.074039.
\bibitem{Y4260_m_6}
T. W. Chiu and T. H. Hsieh, (TWQCD Collaboration),
% Y(4260) on the lattice,
\href{https://doi.org/10.1103/PhysRevD.73.094510}
{Phys.\ Rev.\ D\ {\bf 73}, 094510 (2006).}
%arXiv:hep-lat/0512029,doi:10.1103/PhysRevD.73.094510.
\bibitem{Y4260_nonres1}
E. van Beveren and G. Rupp,
%Interference eects in the X(4260) signal,
\href{https://doi.org/10.1103/PhysRevD.79.111501}
{Phys.\ Rev.\ D\ {\bf 79}, 111501 (2009).}
% arXiv:0905.1595, doi:10.1103/PhysRevD.79.111501.
\bibitem{Y4260_nonres2}
E. van Beveren, G. Rupp, and J. Segovia,
%A Very broad X(4260) and the resonance parameters of the (3D) vector charmonium state,
\href{https://doi.org/10.1103/PhysRevLett.105.102001}
{Phys.\ Rev.\ Lett.\ {\bf 105}, 102001 (2010).}
%arXiv:1005.1010, doi:10.1103/PhysRevLett.105.102001.
\bibitem{Y4260_nonres3}
D. Y. Chen, J. He, and X. Liu,
%Nonresonant explanation for the Y(4260) structure observed in the e+e?? ! J= +?? process,
\href{https://doi.org/10.1103/PhysRevD.83.054021}
{Phys.\ Rev.\ D\ {\bf 83}, 054021 (2011).}
% arXiv:1012.5362, doi:10.1103/PhysRevD.83.054021.
%  %309 citations counted in INSPIRE as of 31 Aug 2016


\bibitem{PhysRevLett.110.252001}
  M.~Ablikim {\it et al.} (BESIII Collaboration),
  %``Observation of a Charged Charmoniumlike Structure in $e^+e^- to pi+ pi- Jpsi$ ''
  \href{https://doi.org/10.1103/PhysRevLett.110.252001}
  {Phys.\ Rev.\ Lett.\  {\bf 110}, 252001 (2013).}
  %doi:10.1103/PhysRevLett.110.252001
  %[arXiv:1303.5949 [hep-ex]].
  %%CITATION = doi:10.1103/PhysRevLett.110.252001;%%
  %398 citations counted in INSPIRE as of 31 Aug 2016
%\cite{PhysRevLett.115.112003}



%33
%\cite{PhysRevLett.112.022001}
\bibitem{PhysRevLett.112.022001}
  M.~Ablikim {\it et al.} (BESIII Collaboration),
  %``Observation of a charged $(D\bar{D}^{*})^\pm$ mass peak in $e^{+}e^{-} \to \pi D\bar{D}^{*}$ at $\sqrt{s} =$ 4.26 GeV,''
\href{https://doi.org/10.1103/PhysRevLett.112.022001}
{Phys.\ Rev.\ Lett.\  {\bf 112}, 022001 (2014).}
  %doi:10.1103/PhysRevLett.112.022001
  %[arXiv:1310.1163 [hep-ex]].
  %%CITATION = doi:10.1103/PhysRevLett.112.022001;%%
  %145 citations counted in INSPIRE as of 31 Aug 2016


%34
%\cite{PhysRev.D92.092006}
\bibitem{PhysRev.D92.092006}
  M.~Ablikim {\it et al.} (BESIII Collaboration),
  %``Confirmation of a charged charmoniumlike state $Z_c(3885)^{\mp}$ in $e^+e^-\to\pi^{\pm}(D\bar{D}^*)^\mp$ with double $D$ tag,''
  \href{https://doi.org/10.1103/PhysRevD.92.092006}
  {Phys.\ Rev.\ D {\bf 92}, 092006 (2015).}
  %doi:10.1103/PhysRevD.92.092006
  %[arXiv:1509.01398 [hep-ex]].
  %%CITATION = doi:10.1103/PhysRevD.92.092006;%%
  %14 citations counted in INSPIRE as of 31 Aug 2016


%35
%\cite{PhysRevLett.115.222002}
\bibitem{PhysRevLett.115.222002}
  M.~Ablikim {\it et al.} (BESIII Collaboration),
  %``Observation of a Neutral Structure near the $D\bar{D}^{*}$ Mass Threshold in $e^{+}e^{-}\to (D \bar{D}^*)^0\pi^0$ at $\sqrt{s}$ = 4.226 and 4.257 GeV,''
  \href{https://doi.org/10.1103/PhysRevLett.115.222002}
  {Phys.\ Rev.\ Lett.\  {\bf 115}, 222002 (2015).}
  %doi:10.1103/PhysRevLett.115.222002
  %[arXiv:1509.05620 [hep-ex]].
  %%CITATION = doi:10.1103/PhysRevLett.115.222002;%%
  %11 citations counted in INSPIRE as of 31 Aug 2016


%36
%\cite{PhysRevLett.112.132001}
\bibitem{PhysRevLett.112.132001}
  M.~Ablikim {\it et al.} (BESIII Collaboration),
  %``Observation of a charged charmoniumlike structure in $e^+e^- \to (D^{*} \bar{D}^{*})^{\pm} \pi^\mp$ at $\sqrt{s}=4.26$GeV,''
  \href{https://doi.org/10.1103/PhysRevLett.112.132001}
  {Phys.\ Rev.\ Lett.\  {\bf 112}, 132001 (2014).}
  %doi:10.1103/PhysRevLett.112.132001
  %[arXiv:1308.2760 [hep-ex]].
  %%CITATION = doi:10.1103/PhysRevLett.112.132001;%%
  %183 citations counted in INSPIRE as of 31 Aug 2016


%37
%\cite{PhysRevLett.115.182002}
\bibitem{PhysRevLett.115.182002}
  M.~Ablikim {\it et al.} (BESIII Collaboration),
  %``Observation of a neutral charmoniumlike state $Z_c(4025)^0$ in $e^{+} e^{-} \to (D^{*} \bar{D}^{*})^{0} \pi^0$,''
  \href{https://doi.org/10.1103/PhysRevLett.115.182002}
  {Phys.\ Rev.\ Lett.\  {\bf 115}, 182002 (2015).}
  %doi:10.1103/PhysRevLett.115.182002
  %[arXiv:1507.02404 [hep-ex]].
  %%CITATION = doi:10.1103/PhysRevLett.115.182002;%%
  %18 citations counted in INSPIRE as of 31 Aug 2016
%%14
%\cite{PhysRevLett.111.242001}
\bibitem{PhysRevLett.111.242001}
  M.~Ablikim {\it et al.} (BESIII Collaboration),
  %``Observation of a Charged Charmoniumlike Structure $Z_c$(4020) and Search for the $Z_c$(3900) in $e^+e^- \to pi+ pi- h_c$,''
  \href{https://doi.org/10.1103/PhysRevLett.111.242001}
  {Phys.\ Rev.\ Lett.\  {\bf 111}, 242001 (2013).}
  %doi:10.1103/PhysRevLett.111.242001
  %[arXiv:1309.1896 [hep-ex]].
  %%CITATION = doi:10.1103/PhysRevLett.111.242001;%%
  %194 citations counted in INSPIRE as of 31 Aug 2016

%
%%15
%\cite{PhysRevLett.113.212002}
\bibitem{PhysRevLett.113.212002}
  M.~Ablikim {\it et al.} (BESIII Collaboration),
  %``Observation of $e^+e^- pi0 pi0 h_c$ and a Neutral Charmoniumlike Structure $Z_c(4020)^0$,''
  \href{https://doi.org/10.1103/PhysRevLett.113.212002}
  {Phys.\ Rev.\ Lett.\  {\bf 113}, 212002 (2014).}
  %doi:10.1103/PhysRevLett.113.212002
  %[arXiv:1409.6577 [hep-ex]].
  %%CITATION = doi:10.1103/PhysRevLett.113.212002;%%
  %50 citations counted in INSPIRE as of 31 Aug 2016
%22
\bibitem{Collaboration:2017njt}
  M.~Ablikim {\it et al.} (BESIII Collaboration),
  %``Determination of the Spin and Parity of the $Z_c(3900)$,''
  \href{https://doi.org/10.1103/PhysRevLett.119.072001}
  {Phys.\ Rev.\ Lett.\  {\bf 119}, 072001 (2017).}
  %doi:10.1103/PhysRevLett.119.072001
  %[arXiv:1706.04100 [hep-ex]].
  %%CITATION = doi:10.1103/PhysRevLett.119.072001;%%
  %3 citations counted in INSPIRE as of 28 Sep 2017



%26
\bibitem{ddmolecule}
F.\ K. Guo, C. Hidalgo-Duque, J. Nieves, and M. P.
Valderrama,
%Consequences of heavy quark symmetries
%for hadronic molecules,
\href{https://doi.org/10.1103/PhysRevD.88.054007}
{Phys.\ Rev.\ D {\bf 88}, 054007 (2013).}

%27
\bibitem{tetraquark}
L. Maiani, V. Riquer, R. Faccini, F. Piccinini, A. Pilloni,
and A. D. Polosa,
%A JPG . 1tt charged resonance in the
%Yd4260T ¨ ƒÎtƒÎ.J=ƒÕ decay?,
\href{https://doi.org/10.1103/PhysRevD.87.111102}
{Phys.\ Rev.\ D {\bf 87}, 111102 (2013).}

%28
\bibitem{cusps}
E. S. Swanson,
%Coupled channel cusps and Zbe10610T,Zbe10650T, and Zce3900T,
\href{https://doi.org/10.1103/PhysRevD.91.034009}
{Phys.\ Rev.\ D {\bf 91}, 034009 (2015).}

%29
\bibitem{threshold1}
Q. Wang, C. Hanhart, and Q. Zhao,
%Decoding the Riddle of Ye4260T and Zce3900T,
\href{https://doi.org/10.1103/PhysRevLett.111.132003}
{Phys.\ Rev.\ Lett. {\bf 111}, 132003 (2013).}

%30
\bibitem{threshold2}
D. Y. Chen, X. Liu, and T. Matsuki,
%Predictions of Charged Charmoniumlike Structures with Hidden-Charm and Open-Strange Channels,
\href{https://doi.org/10.1103/PhysRevLett.110.232001}
{Phys.\ Rev.\ Lett. {\bf 110}, 232001 (2013).}


%21
\bibitem{PhysRev.D92.032009}
  M.~Ablikim {\it et al.} (BESIII Collaboration),
  %``Search for $Z_c(3900)^\pm\to\omega\pi^\pm$,''
  \href{https://doi.org/10.1103/PhysRevD.92.032009}
  {Phys.\ Rev.\ D {\bf 92}, 032009 (2015).}
  %doi:10.1103/PhysRevD.92.032009
  %[arXiv:1507.02068 [hep-ex]].
  %%CITATION = doi:10.1103/PhysRevD.92.032009;%%
  %6 citations counted in INSPIRE as of 31 Aug 2016






%24
\bibitem{lhd1}
M.~Ablikim {\it et al.} (BESIII Collaboration),
%Measurements of cross section of e+e?¡úpp¡¥¦Ð0 at center-of-mass energies between 4.008 and 4.600 GeV,
\href{https://doi.org/10.1016/j.physletb.2017.05.033}
{Phys.\ Lett.\ B {\bf 771}, 45 (2017).}
%ISSN 0370-2693, %https://doi.org/10.1016/j.physletb.2017.05.033.
%(http://www.sciencedirect.com/science/article/pii/S0370269317303969)
%Keywords: Hadrons; Cross section measurements; Y(4260)


%25
\bibitem{lhd2}
M.~Ablikim {\it et al.} (BESIII Collaboration),
\href{https://doi.org/10.1016/j.physletb.2017.09.021}
{Phys.\ Lett.\ B {\bf 774}, 78 (2017).}

\bibitem{PDG}
M. Tanabashi et al. (Particle Data Group),
\href{https://doi.org/10.1103/PhysRevD.98.030001}
{Phys. Rev. D {\bf 98}, 030001 (2018).}


%40
\bibitem{Ablikim2010345} M. Ablikim {\it et al.} (BESIII Collaboration),
\href{https://doi.org/10.1016/j.nima.2009.12.050}
{Nucl. Instrum. Meth. A {\bf 614}, 345 (2010).}
%41
\bibitem{BEPCII} D. M. Asner {\it et al.},
\href{https://arxiv.org/abs/0809.1869}
{Int. J. Mod. Phys. A {\bf 24}, S1 (2009).}
%, arXiv:0809.1869.% [hep-ex].
%42
\bibitem{Ablikim:2015nan}
  M.~Ablikim {\it et al.} (BESIII Collaboration),
  %``Precision measurement of the integrated luminosity of the data taken by BESIII at center of mass energies between 3.810 GeV and 4.600 GeV,''
  \href{https://doi.org/10.1088/1674-1137/39/9/093001}
  {Chin.\ Phys.\ C {\bf 39}, 093001 (2015).}
  %doi:10.1088/1674-1137/39/9/093001
  %[arXiv:1503.03408 [hep-ex]].
  %%CITATION = doi:10.1088/1674-1137/39/9/093001;%%
  %48 citations counted in INSPIRE as of 26 Sep 2017
%43
\bibitem{Ablikim:2015zaa}
  M.~Ablikim {\it et al.} (BESIII Collaboration),
  %``Measurement of the center-of-mass energies at BESIII via the di-muon process,''
  \href{https://doi.org/10.1088/1674-1137/40/6/063001}
  {Chin.\ Phys.\ C {\bf 40}, 063001 (2016).}
  %doi:10.1088/1674-1137/40/6/063001
  %[arXiv:1510.08654 [hep-ex]].
  %%CITATION = doi:10.1088/1674-1137/40/6/063001;%%
  %17 citations counted in INSPIRE as of 26 Sep 2017

%44
\bibitem{geant4} S.~Agostinelli {\it et al.} ({\sc geant4} Collaboration),
\href{https://doi.org/10.1016/S0168-9002(03)01368-8}
{Nucl.\ Instrum.\ Meth.\ A {\bf 506}, 250 (2003).}
%45
\bibitem{ref:boost} Z. Y. Deng {\it et al.},
\href{http://cpc-hepnp.ihep.ac.cn:8080/Jwk_cpc/EN/Y2006/V30/I05/371}
{Chin. Phys. C {\bf 30}, 371 (2006).}
%46

%47
\bibitem{EVTGEN2} D.~J.~Lange,
\href{https://doi.org/10.1016/S0168-9002(01)00089-4}
{Nucl.\ Instrum.\ Meth.\ A {\bf 462}, 152 (2001).}
\bibitem{EVTGEN} R. G. Ping,
%\href{http://cpc-hepnp.ihep.ac.cn:8080/Jwk_cpc/EN/Y2008/V32/I08/599}
\href{https://doi.org/10.1088/1674-1137/32/8/001}
{Chin. Phys. C {\bf 32}, 599 (2008).}
%51
\bibitem{KKMC} S. Jadach, B. F. L. Ward, and Z. Was,
\href{https://doi.org/10.1103/PhysRevD.63.113009}
{Phys. Rev. D {\bf 63}, 113009 (2001).}
%52
\bibitem{Babayaga}
G. Balossini, C. M. C. Calame, G. Montagna,
O. Nicrosini, and F. Piccinini.
\href{https://doi.org/10.1016/j.nuclphysb.2006.09.022}
{Nucl. Phys. B {\bf 758}, 227 (2006).}

\bibitem{SecondVFit}  M. Xu {\it et al.},
\href{https://doi.org/10.1088/1674-1137/33/6/005}
{Chin. Phys. C {\bf 33}, 428 (2009).}
\bibitem{VP}
S. Actis \emph{et al.},
%(Working Group on Radiative Corrections and Monte Carlo Generators for Low Energies Collaboration), Quest for precision in hadronic cross sections at low energy:
%Monte Carlo tools vs. experimental data,
\href{https://doi.org/10.1140/epjc/s10052-010-1251-4}
{Eur. Phys. J. C {\bf 66}, 585 (2010).}
%53
\bibitem{Kuraev:1985hb}
  E.~A.~Kuraev and V.~S.~Fadin,
  %``On Radiative Corrections to e+ e- Single Photon Annihilation at High-Energy,''
  \href{http://inspirehep.net/record/217313?ln=zh_CN}
  {Sov.\ J.\ Nucl.\ Phys.\  {\bf 41}, 466 (1985).}
  %[Yad.\ Fiz.\  {\bf 41}, 733 (1985)].
  %%CITATION = SJNCA,41,466;%%
  %751 citations counted in INSPIRE as of 26 Sep 2017
%54



%55
\bibitem{estimator} X. H. Mo,
\href{http://cpc-hepnp.ihep.ac.cn:8080/Jwk_cpc/EN/Y2007/V31/I08/745}
{HEP \& NP. {\bf 31}, 745 (2007).}
%56
%\cite{PhysRevLett.110.252001}
\bibitem{PhysRevLett.115.112003}
  M.~Ablikim {\it et al.} (BESIII Collaboration),
  %``Observation of $Z_c(3900)^{0}$ in $e^+e^-\to\pi^0\pi^0 J/\psi$,''
  \href{https://doi.org/10.1103/PhysRevLett.115.112003}
  {Phys.\ Rev.\ Lett.\  {\bf 115}, 112003 (2015).}
  %doi:10.1103/PhysRevLett.115.112003
  %[arXiv:1506.06018 [hep-ex]].
  %%CITATION = doi:10.1103/PhysRevLett.115.112003;%%
  %31 citations counted in INSPIRE as of 31 Aug 2016


%57
\bibitem{Ablikim:2011kv}
  M.~Ablikim \emph{et al.} (BESIII Collaboration),
  %``Study of $\chi_{cJ}$ radiative decays into a vector meson,''
  \href{https://doi.org/10.1103/PhysRevD.83.112005}
  {Phys.\ Rev.\ D {\bf 83}, 112005 (2011).}
  %doi:10.1103/PhysRevD.83.112005
  %[arXiv:1103.5564 [hep-ex]].
  %%CITATION = doi:10.1103/PhysRevD.83.112005;%%
  %71 citations counted in INSPIRE as of 29 Sep 2017

\bibitem{pi0_rec_eff}
  M.~Ablikim \emph{et al.} (BESIII Collaboration),
  %``Branching fraction measurements of \chi_{c0} and \chi_{c2} to \pi^0\pi^0 and \eta\eta,''
  \href{https://doi.org/10.1103/PhysRevD.81.052005}
  {Phys.\ Rev.\ D {\bf 81}, 052005 (2010).}
  %doi:10.1103/PhysRevD.81.052005
  %[arXiv:1001.5360 [hep-ex]].
  %%CITATION = doi:10.1103/PhysRevD.81.052005;%%
  %77 citations counted in INSPIRE as of 29 Sep 2017

\bibitem{Ks_err} M. Ablikim {\it et al.} (BESIII Collaboration),
\href{https://doi.org/10.1103/PhysRevD.92.112008}
{Phys.\ Rev.\ D. {\bf 92}, 112008 (2015).}
%59
\bibitem{Ablikim:2012pg}
  M.~Ablikim {\it et al.} (BESIII Collaboration),
  %``Search for hadronic transition $¦Ö_{cJ} ¡ú ¦Ç_c¦Ð^+¦Ð^-$ and observation of $¦Ö_{cJ} ¡ú K\overline{K}¦Ð¦Ð¦Ð$,''
  \href{https://doi.org/10.1103/PhysRevD.87.012002}
  {Phys.\ Rev.\ D {\bf 87}, 012002 (2013).}
  %doi:10.1103/PhysRevD.87.012002
  %[arXiv:1208.4805 [hep-ex]].
  %%CITATION = doi:10.1103/PhysRevD.87.012002;%%
  %45 citations counted in INSPIRE as of 29 Sep 2017
\bibitem{smear}
K. Stenson,
\href{https://arxiv.org/abs/physics/0605236}{arXiv:physics/0605236}.
\bibitem{Ablikim:2016qzw}
  M.~Ablikim {\it et al.} (BESIII Collaboration),
  %``Precise measurement of the $e^+e^-\to \pi^+\pi^-J/\psi$ cross section at center-of-mass energies from 3.77 to 4.60 GeV,''
  \href{https://doi.org/10.1103/PhysRevLett.118.092001}
  {Phys.\ Rev.\ Lett.\  {\bf 118}, 092001 (2017).}
%  doi:10.1103/PhysRevLett.118.092001
%  [arXiv:1611.01317 [hep-ex]].
  %%CITATION = doi:10.1103/PhysRevLett.118.092001;%%
  %23 citations counted in INSPIRE as of 28 Sep 2017

\end{thebibliography}
\end{document}